\documentclass[twocolumn]{aastex6}

\usepackage{amsmath}

\newcommand{\keyw}[1]{\textcolor{gray}{#1}}
\newcommand{\dt}[1]{\textcolor{black}{#1}}
\begin{document}

\title{A Millimeter Continuum Size--Luminosity Relationship for Protoplanetary Disks}
\author{Anjali Tripathi\altaffilmark{1}, Sean M.~Andrews\altaffilmark{1}, Tilman Birnstiel\altaffilmark{2}, \& David J.~Wilner\altaffilmark{1}}
\altaffiltext{1}{Harvard-Smithsonian Center for Astrophysics, 60 Garden Street, Cambridge, MA 02138, USA; \url{atripathi}, \url{sandrews@cfa.harvard.edu}}
\altaffiltext{2}{University Observatory, Faculty of Physics, Ludwig-Maximilians-Universit\"at M\"unchen, Scheinerstr.~1, 81679 Munich, Germany}

\begin{abstract}
We present a \dt{sub-arcsecond} resolution survey of the 340\,GHz dust continuum emission from 50 nearby protoplanetary disks, based on new and archival observations with the Submillimeter Array.  The observed visibility data were modeled with a simple prescription for the radial surface brightness profile.  The results were used to extract intuitive, empirical estimates of the emission ``size" for each disk, $R_{\rm eff}$, defined as the radius that encircles a fixed fraction of the total continuum luminosity, $L_{\rm mm}$.  We find a significant correlation between the sizes and luminosities, such that $R_{\rm eff} \propto L_{\rm mm}^{0.5}$, providing a confirmation and quantitative characterization of a putative trend that was noted previously.  This correlation suggests that these disks have roughly the same {\it average} surface brightness interior to their given effective radius, $\sim$0.2\,Jy\,arcsec$^{-2}$ (or 8\,K in brightness temperature).  The same trend remains, but the 0.2\,dex of dispersion perpendicular to this relation essentially disappears, when we account for the irradiation environment of each disk with a crude approximation of the dust temperatures based on the stellar host luminosities.  We consider two (not mutually exclusive) explanations for the origin of this size--luminosity relationship.  Simple models of the growth and migration of disk solids can account for the observed trend for a reasonable range of initial conditions, but only on timescales that are much shorter than the nominal ages present in the sample.  An alternative scenario invokes optically thick emission concentrated on unresolved scales, with filling factors of a few tens of percent, that are perhaps manifestations of localized particle traps.           
\end{abstract}
\keywords{\keyw{circumstellar matter --- planetary systems: formation, protoplanetary disks --- dust}}

\section{Introduction}

Observations that help characterize the relationships among key properties of protoplanetary disks, and those of their host stars, are essential to developing a more robust and complete planet formation theory.  Measurements of continuum emission at (sub-)millimeter wavelengths (hereafter ``mm") are important since they trace the reservoir of planetesimal precursors.  They are especially sensitive diagnostics of the growth and migration of solids during the early stages of planet formation.  

Despite the complexity of the planet formation process, theoretical models agree that overall efficiency depends strongly on the mass of raw material available in the disk.  The most accessible diagnostic of mass is the mm continuum luminosity ($L_{\rm mm}$) generated by dust grains \citep{beckwith90}.  Because of its presumed low optical depth, $L_{\rm mm}$ scales roughly linearly with the total mass and mean temperature of the disk solids.  With reasonable assumptions for the grain properties, mm luminosity surveys have been used to construct disk mass distributions \citep{aw05,aw07a}, and study how they vary with time \citep{mathews12,williams13,carpenter14,ansdell15,cieza15,barenfeld16}, environment \citep{jensen94,jensen96,harris12,akeson14,mann09,mann10,mann14}, and stellar host mass \citep{andrews13,mohanty13,carpenter14,ansdell16,barenfeld16,pascucci16}. 

By itself, an {\it unresolved} quantity like $L_{\rm mm}$ provides little leverage for constraining the wide diversity of disk properties relevant to planet formation.  Information on the spatial distribution of the disk material is needed for a more robust characterization of the disk population.  With even crude resolution, another elementary disk property -- {\it size} -- becomes accessible.  Disk sizes reflect some convolution of the processes involved in formation \citep{galli06,mellon08,vorobyov09,dapp10,machida11}, evolution of angular momentum \citep[e.g.,][]{lyndenbell74,hartmann98}, and, perhaps most importantly, the growth and transport of the constituent solid particles \citep[e.g.,][]{birnstiel14}.  

\begin{figure*}[t!]
\includegraphics[width=\linewidth]{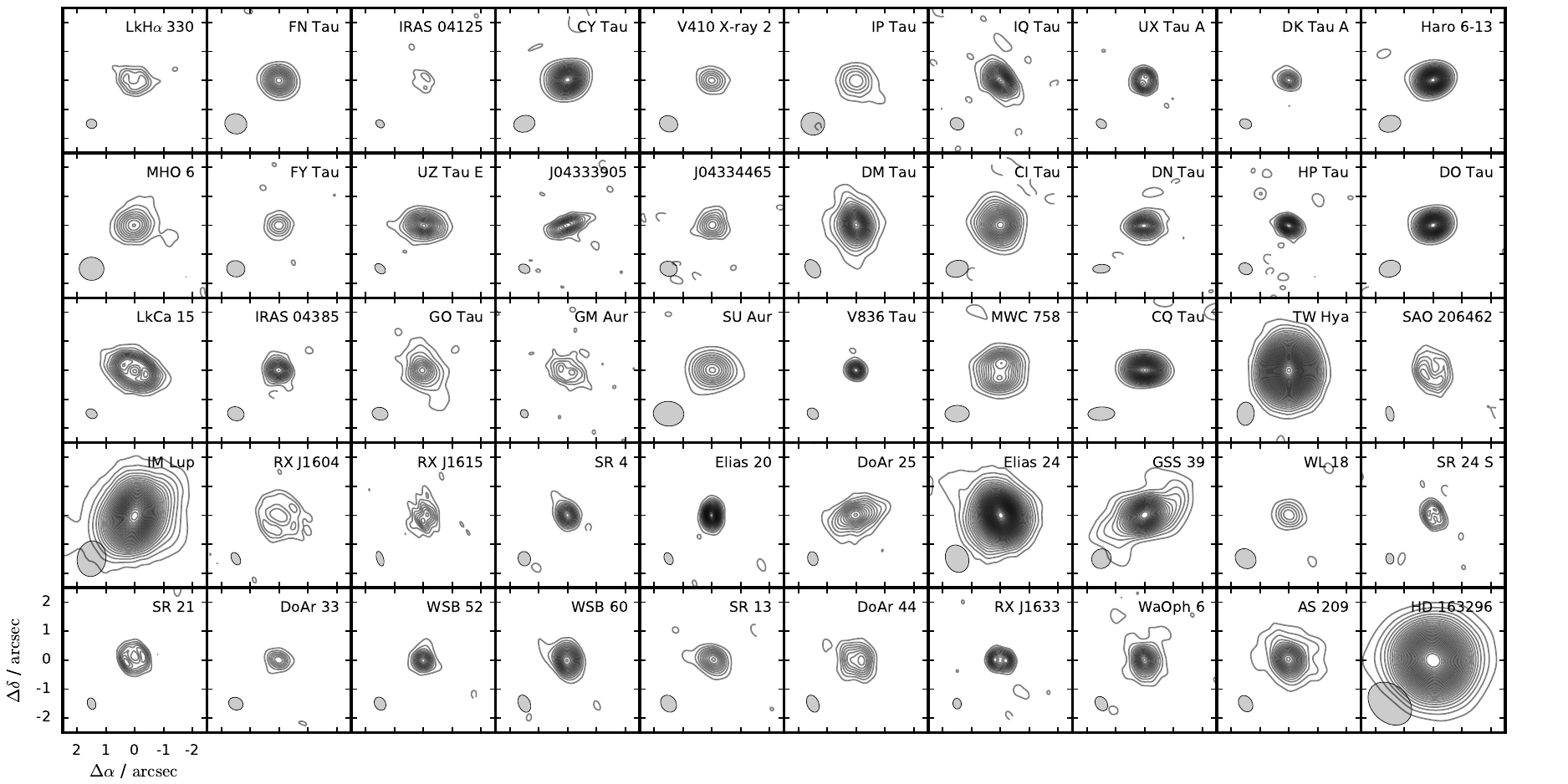}
\figcaption{A gallery of 340\,GHz continuum images for the entire sample.  Synthesized beam dimensions are marked in the lower left corner of each panel.  Contours are drawn at intervals of 3$\times$ the RMS noise level.  Basic imaging parameters are provided in Table~\ref{table:images}.
\label{fig:gallery}}
\end{figure*}

Work in this field has been relatively restricted to detailed case studies, but nevertheless a compelling demographic trend related to sizes and masses has already been identified.  Based on a small 340\,GHz sample, \citet{andrews10b} found that disks with lower luminosities (i.e., less massive) are preferentially smaller, but do not necessarily have lower surface brightnesses.  \citet{pietu14} confirmed this tendency, with a different sample and at a slightly lower frequency (230\,GHz). 

Crucial information about the mechanisms at play in planetesimal formation and transport in disks are potentially encoded in this putative {\it size--luminosity relationship}.  Our goal is to validate and better characterize this trend.  With that motivation in mind, we present a substantially larger survey of resolved protoplanetary disk continuum measurements that spans a wide range of the relevant parameter-space, in an effort to better characterize how disk sizes are related to their luminosities.  We introduce this dataset in Section \ref{sec:data} and describe an empirical, intuitive methodology for inferring a robust size metric for each disk in Section \ref{sec:model}.  The results of this homogenized analysis are presented in Section \ref{sec:results} and interpreted in the context of our current understanding of the evolution of disk solids in Section \ref{sec:discussion}.

\section{Data} \label{sec:data}

\subsection{Sample}

A sample of 50 nearby ($d\le200$\,pc) disk targets was collated from the archived catalog of $\sim$340\,GHz (880\,$\mu$m) continuum measurements made with the Submillimeter Array \citep[SMA;][]{ho04}, since the 2004 start of science operations.  We considered four primary selection criteria to include a given target in this sample: (1) a quality calibration (i.e., suitable phase stability); (2) a 340\,GHz continuum flux density $\ge$20\,mJy; (3) baseline lengths $\ge$200\,m (to ensure sufficient resolution for estimating an emission size); and (4) no known companion within a $\sim$2\arcsec\ separation.\footnote{An exception is made for very close spectroscopic binary companions (e.g., UZ Tau E), where the dust generating the emission is confined to a circumbinary disk around both components.}  The first three of these are practical restrictions to ensure reliable measurements of the metrics of interest.  The fourth criterion has a physical motivation: it was designed to avoid targets where dynamical interactions in multiple star systems are known to reduce the continuum sizes and luminosities \citep[e.g.,][]{jensen96,harris12}.

Of the 50 disks in our survey, 10 of these were recently observed by us expressly for the purposes of the present study.  To our knowledge, the SMA observations of 18 targets have not yet been published elsewhere.  The data for the remaining 32 targets have appeared previously in the literature (Sect.~\ref{sect:olddata}).  Table~\ref{table:obslog} is a brief SMA observation log, with references for where the data originally appeared. Targets in this sample are primarily located in the Taurus and Ophiuchus star-forming regions, although 9 are in other regions or are found in isolation.

\subsection{New and Previously Unpublished Data\label{sect:newdata}}

We observed 10 disks in the Taurus-Auriga complex (DN Tau, FY Tau, GO Tau, HP Tau, IP Tau, IQ Tau, IRAS 04385+2550, 2MASS J04333905+2227207, 2MASS J04334465+2615005, and MHO 6) using the eight 6-m SMA antennas arranged in their compact, extended, \dt{and very extended} configurations (9--\dt{508}\,m baseline lengths).  The dual-sideband 345\,GHz receivers were tuned to a local oscillator (LO) frequency of 341.6\,GHz (878\,$\mu$m).  The SMA correlator processed two intermediate frequency (IF) bands, at $\pm$4--6 and $\pm$6--8\,GHz from the LO.  Each IF band contains 24 spectral chunks of 104\,MHz width.  One chunk in the lower IF was split into 256 channels; all others were divided into 32 coarser channels.  Each observation cycled between multiple targets and the nearest bright quasars, 3C\,111 and J0510+180.  Additional measurements of 3C\,454.3, Uranus, \dt{and Callisto} were made for bandpass and flux calibration, respectively.

Another 8 disks (CI Tau, CY Tau, DO Tau, Haro 6-13, SU Aur, V410 X-ray 2, V836 Tau, and IM Lup) had previously unpublished data that are suitable for this program available in the SMA archive.  These observations have slight variations on the correlator setup and calibrator sources noted 
above, but in the end produce effectively similar continuum data products.  

The raw visibility data were calibrated using standard procedures with the facility {\tt MIR} package.  After calibrating the spectral response of the system and setting the amplitude scale, gain variations were corrected based on repeat observations of the nearest quasar.  For observations taken over long time baselines, the visibility phases were shifted to align the data: usually this is based on the known proper motion of the stellar host, but in a few cases where these were unavailable we relied on Gaussian fits to individual measurements.  All visibility data for a given target were combined, after confirming their consistency on overlapping baselines.  Each visibility set was Fourier inverted, deconvolved with the {\tt clean} algorithm, and then convolved with a restoring beam to synthesize an image.  The imaging process was conducted with the {\tt MIRIAD} software package.  Table~\ref{table:images} lists basic imaging parameters for the composite datasets.  Figure~\ref{fig:gallery} shows a gallery of images for all targets in the sample.

\subsection{Supplementary Archival Data\label{sect:olddata}}

The bulk of the sample, 32 targets, have SMA continuum data that were already published in the literature.  These disks are primarily in the Ophiuchus star-forming region, but also include members of the Taurus and Lupus complexes, as well as assorted isolated sources.  The details of these observations and their calibration are provided in the original references (see Table~\ref{table:obslog}), but they are generally similar to those presented in Section~\ref{sect:newdata}.

\section{Modeling the Resolved Emission} \label{sec:model}

\subsection{Surface Brightness Model Definition}

High quality observations of disk continuum emission \dt{have previously been modeled with ``broken power-law"  \citep[e.g.,][]{andrews12,degregorio-monsalvo13,hogerheijde16} or ``similarity solution" surface brightness profiles \citep[e.g.,][]{andrews09,andrews10b,isella09,isella10}.  For the survey data presented here, experimentation with these brightness profiles demonstrated that some targets were much better described by one of these options, while the other left significant, persistent residuals.  Instead of using different models for different targets, we elected to adopt a more flexible prescription to interpret the data for the full sample in a homogeneous context.}

\dt{The required flexibility in this prescription is a mechanism that treats either smooth or sharp transitions in the brightness profile.}  Fortunately, analogous studies of elliptical galaxy brightness profiles provide some useful guidance in this context.  \citet{lauer95} introduced a prescription (the so-called ``Nuker" profile) that is \dt{mathematically well-suited to this task},      
\begin{equation}
    I_{\nu}(\varrho) \propto \left( \frac{\varrho}{\varrho_t} \right)^{-\gamma} \left[ 1 + \left (\frac{\varrho}{\varrho_t} \right) ^{\alpha} \right ] ^{(\gamma - \beta)/\alpha},
    \label{eq:nuker}
\end{equation}
where $\varrho$ is the radial coordinate projected on the sky. The Nuker profile has five free parameters: (1) a transition radius $\varrho_t$, (2) an inner disk index $\gamma$, (3) an outer disk index $\beta$, (4) a transition index $\alpha$, and (5) a normalization constant, which we re-cast to be the total flux density $F_\nu$ (defined so that $F_\nu = 2\pi \int I_{\nu}(\varrho) \varrho \, d\varrho$).  

\begin{figure}[t!]
\includegraphics[width=\linewidth]{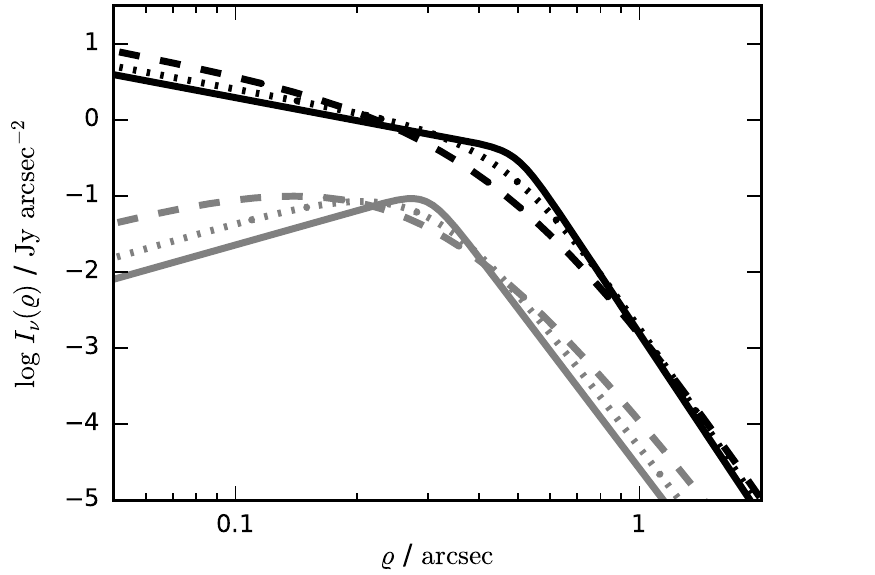}
\figcaption{Representative Nuker profiles.  The black and gray sets of curves show different model families, each of which has the same \{$F_\nu$, $\varrho_t$, $\beta$, $\gamma$\} (\{0.7\,Jy, 0\farcs5, 8, 1\} in black; \{0.3\,Jy, 0\farcs3, 7, -1.5\} in gray), but different $\alpha$: 2 (dashed), 4 (dotted), and 16 (solid).  As $\alpha$ increases, the profiles converge to a sharp broken power-law.  
\label{fig:nuker}}
\end{figure}

It is instructive to consider some asymptotic behavior.  When $\varrho \ll \varrho_t$ or $\varrho \gg \varrho_t$, the brightness profile scales like $\varrho^{-\gamma}$ or $\varrho^{-\beta}$, respectively.  The index $\alpha$ controls where these asymptotic behaviors are relevant: a higher $\alpha$ value pushes the profile toward a sharp broken power-law morphology.  The same parameterization can reproduce the emission profiles for standard, continuous disk models as well as the ring-like emission noted for ``transition" disks \citep[e.g.,][]{andrews11}.  Figure~\ref{fig:nuker} shows Nuker profiles for some representative parameter values.  

From a given Nuker profile, we compute a set of model visibilities from the Fourier transform of Eq.~(\ref{eq:nuker}), sampled at the same discrete set of spatial frequencies as the data.  Those visibilities are modified (stretched, rotated, and shifted) to account for four geometric parameters relevant to the observations: (1) a projected inclination angle $i$, (2) a rotation of the major axis in the plane of the sky $\varphi$ (position angle), and (3)+(4) position offsets from the observed phase center \{$d\alpha$, $d\delta$\}.  Taken together, nine parameters, $\theta$ = [$\varrho_t$, $\gamma$, $\beta$, $\alpha$, $F_{\nu}$, $i$, $\varphi$, $d\alpha$, $d\delta$], fully specify a set of model visibilities, $\mathcal{V}_{\rm m}(\theta)$.

\dt{It is worth highlighting two important points about the choice of this model prescription.  First, the specific parameterization adopted for the brightness profile does not matter in the specific context of this study, {\it so long as it accurately describes the data} (i.e., leaves no statistically significant residuals).  This will be illustrated more directly below, after a description of the full modeling procedure.  Second, we have expressly avoided casting the interpretation in a physical context.  Converting the results into inferences on optical depths or temperatures implicitly introduces a set of strong degeneracies and a model dependence that is not well-motivated from a physical standpoint.  Our aim is instead to hew as close to the observations as possible; we will consider some connections to physical properties only in the larger context of the results from the full sample (see Sect.~\ref{sec:discussion}).}

\subsection{Surface Brightness Inference} \label{sec:SB}

To compare a given model to the visibility data, $\mathcal{V}_{\rm d}$, we employ a standard Gaussian likelihood,
\begin{equation}
\log{p(\mathcal{V}_{\rm d} | \theta)} = -\frac{1}{2} \sum_i w_{{\rm d}, i} | \mathcal{V}_{{\rm d}, i} - \mathcal{V}_{{\rm m}, i}(\theta) |^2,
\end{equation}
the sum of the residual moduli weighted by the standard natural visibility weights.  The classic inference problem can be cast with the posterior probability distribution of the model parameters conditioned on the data as
\begin{equation}
\log{p(\theta | \mathcal{V}_{\rm d})} = \log{p(\mathcal{V}_{\rm d} | \theta)} + \log{p(\theta)} + {\rm constant},
\end{equation}
where $p(\theta) = \prod_j p(\theta_j)$, the product of the priors for each parameter (presuming their independence).

We assign uniform priors for most parameters: $p(F_\nu) = \mathcal{U}(0$, 10\,Jy), $p(\varrho_t) = \mathcal{U}(0$, 10\arcsec), $p(\varphi) = \mathcal{U}(0$, 180\degr), and $p(d\alpha)$, $p(d\delta) = \mathcal{U}(-3$, +3\arcsec).  A simple geometric prior is adopted for the inclination angle, $p(i) = \sin{i}$.  

Priors for the Nuker profile's index parameters are not obvious.  They were assigned based on some iterative experimentation, grounded in our findings for the targets that have data with higher sensitivity and resolution.  We set $p(\beta) = \mathcal{U}(2, 10)$: the low bound forces the intensity profile to decrease at large radii (as is observed), and the high bound is practical -- at these resolutions, the data are not able to differentiate between more extreme indices.  To appropriately span both smooth and sharp transitions between the two power-law regimes, we set $p(\log_{10}{\alpha}) = \mathcal{U}(0, 2)$ (see Fig.~\ref{fig:nuker}).  We sample the posterior in $\log_{10}{\alpha}$ rather than $\alpha$, since most of the diversity in Nuker profile shapes happens in the first decade of the prior-space.  Finally, we set a softer-edged prior on $\gamma$ that is approximately uniform over the range $(-3, 2)$, but employs logistic tapers on the boundaries:
\begin{equation}
p(\gamma) \propto \frac{1}{1+e^{-5(\gamma+3)}} - \frac{1}{1+e^{-15(\gamma-2)}}.
\end{equation}
The low-$\gamma$ bound is again practical (like the high-$\beta$ bound).  The steeper bound at high $\gamma$ values was designed to improve convergence for the poorly-resolved cases: when $\gamma \gtrsim 2$, the degeneracy with $\varrho_t$ is severe and both parameters can increase  without bound (i.e., very steep gradients can be accommodated for very large transition radii).  Since none of the well-resolved targets ended up having $\gamma \gtrsim 1$ (regardless of the $\gamma$ prior), we consider this adopted upper boundary conservative.    

A Markov Chain Monte Carlo (MCMC) algorithm was used to explore the posterior probability space.  We employed the ensemble sampler proposed by \citet{goodman10} and implemented as the open-source code package {\tt emcee} by \citet{foreman-mackey13}.  With this algorithm, we used 48 ``walkers" to sample in $\log{p(\theta | \mathcal{V}_{\rm d})}$ and ran 50,000 steps per walker.

To identify a starting point for parameter estimation, we first crudely estimate the disk geometric parameters \{$i$, $\varphi$, $d\alpha$, $d\delta$\} with an elliptical Gaussian fit to the visibility data.  Using these rough values, we deproject and azimuthally average the visibilities into a one-dimensional profile as a function of baseline length.  This is used as a guide to visually probe models with different Nuker profile parameters, \{$F_\nu$, $\varrho_t$, $\alpha$, $\beta$, $\gamma$\}.  When a reasonable match is found (with some experience, this takes about a minute), we assign a conservative (i.e., broad) range around each of the parameters.  The ensemble sampler walkers are then initialized with random draws from uniform distributions that span those ranges. 

While assessing formal convergence is difficult for this algorithm (since the walkers are co-dependent), we have used a variety of simple rubrics (e.g., trace and autocorrelation examinations, an inter-walker \citealt{gelman92} test, and the comparison of sub-chain means and variances advocated by \citealt{geweke92}) that lend confidence that we have reached a stationary target distribution.  Autocorrelation lengths for all parameters are of the order 10$^2$ steps, implying that (after excising burn-in steps) we have $\gtrsim$10$^4$ independent samples of the posterior for each target.  We find acceptance fractions of 0.2--0.4.

\subsection{Disk Size Metric} \label{sec:size}

\begin{figure}[t!]
\includegraphics[width=\linewidth]{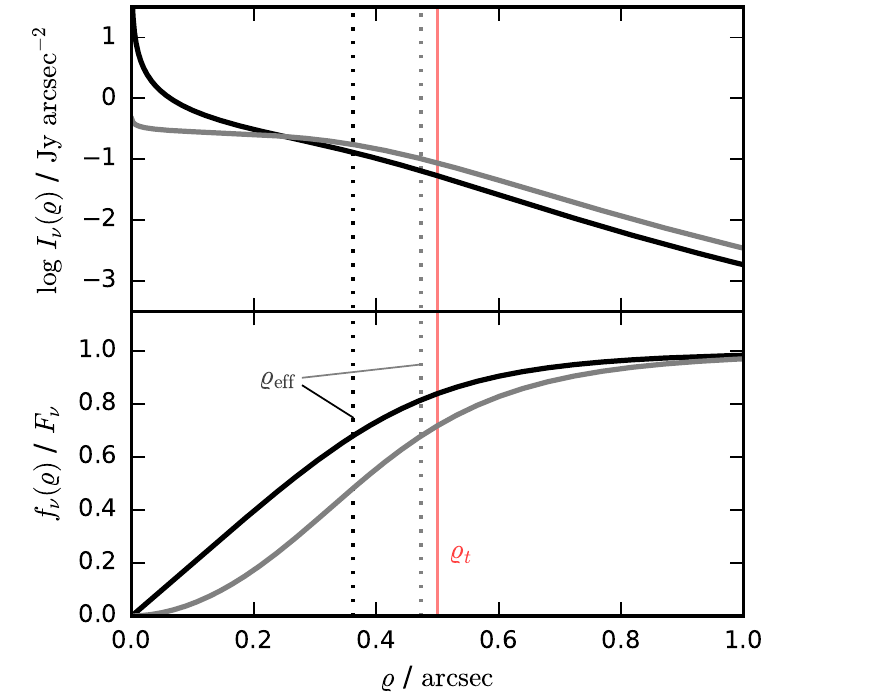}
\figcaption{
Two brightness profiles with different inner disk indices, $\gamma=1$ in black and $\gamma=0.1$ in gray, but otherwise identical parameters ($F_\nu = 0.2$\,Jy, $\varrho_t = 0\farcs5$, $\alpha = 4$, $\beta = 6$).  Despite having the same $\varrho_t$ (red), the model with lower $\gamma$ appears larger to an observer (it has more emission distributed at larger radii).  That fact is better reflected in the {\it effective} size metric we have adopted, $\varrho_{\rm eff}$ (dotted), defined as the location where $f_\nu(\varrho)/F_\nu =0.68$.
\label{fig:sizemeaning}}
\end{figure}

The disk size is {\it not} one of the parameters that we infer directly.  If we followed the traditional methodology, we would proceed by taking $\varrho_t$ as the size metric.  But that approach is problematic; $\varrho_t$ is not really how we think about an emission size.  As an illustration, consider the two disk models shown in Figure~\ref{fig:sizemeaning} with identical $\varrho_t$ but different $\gamma$ values.  Which is ``larger"?   Our instinct suggests the disk with a lower $\gamma$ is larger, since its emission clearly has a more radially extended morphology.  

\dt{If we consider alternative prescriptions for the surface brightness profile, there is an analogous ambiguity.  Any such model will include a ``scale" parameter (like $\varrho_t$) that will generally have a different value than the one inferred from the data for the Nuker profile.  Given the limitations of the data imposed by noise and resolution constraints, we cannot definitively determine which of these profiles, and their corresponding size metrics, is most appropriate.}  As a more practical concern, a degeneracy with the parameters that describe the gradients of the Nuker profile\footnote{This same degeneracy problem is present for {\it any} brightness profile \citep[e.g.,][]{mundy96,aw07a}.}, \{$\alpha$, $\beta$, $\gamma$\}, makes it difficult to obtain a precise inference of $\varrho_t$ using data with typical sensitivity and resolution (especially if $\alpha \lesssim 10$, since the resulting smoother profile makes the transition less distinct). 

There is a generic definition of size that alleviates these issues.  If we construct a {\it cumulative} intensity profile,
\begin{equation}
f_{\nu}(\varrho) = 2 \pi \int_0^{\varrho} I_{\nu}(\varrho^\prime) \, \varrho^\prime \, d\varrho^\prime,
\label{eq:fcum}
\end{equation}
then we can assign an {\it effective radius}, $\varrho_{\rm eff}$, that encircles a fixed fraction, $x$, of the total flux: $f_\nu(\varrho_{\rm eff}) = x F_{\nu}$ for some $x \in [0,1]$ (note that $F_\nu = f_\nu(\infty)$ by definition).  The inference of $\varrho_{\rm eff}$ is considerably less affected by imprecise constraints on the gradient parameters.  Moreover, $\varrho_{\rm eff}$ more faithfully captures the intuitive intent of a size metric, \dt{as shown in Figure \ref{fig:sizemeaning}}.  Most importantly, it has the same straightforward meaning regardless of the underlying surface brightness profile\dt{; any profile that accurately reproduces the data will have the same $\varrho_{\rm eff}$}.

That said, the selection of $x$ in the definition of $\varrho_{\rm eff}$ is technically arbitrary.  It makes sense to fix $x$ and therefore homogenize the analysis for a sample.  But, there are some practical concerns to take into consideration.  If $x$ is too low, then $\varrho_{\rm eff}$ would be uncomfortably reliant on a sub-resolution extrapolation of the Nuker profile.  And if $x$ is too high, then $\varrho_{\rm eff}$ would simply reflect the quality of the constraint on $\beta$ (and in some cases $\alpha$) based on the part of the brightness profile where we have the poorest sensitivity (large $\varrho$).  Some of the more intuitive choices are $x = 0.50$ (so $\varrho_{\rm eff}$ is the ``half-light" radius) or 0.68 (so $\varrho_{\rm eff}$ is comparable to a ``standard deviation" in the admittedly poor approximation of a Gaussian brightness profile).  Given the concern about extrapolation expressed above, we prefer to define $\varrho_{\rm eff}$ based on $x = 0.68$.  An alternative choice in the range $x \in [0.5, 0.8]$ makes little difference in the analysis or results that follow. 

We derive posterior samples for $\varrho_{\rm eff}$ by integrating the cumulative intensity profiles (i.e., solving Eq.~\ref{eq:fcum}) for each brightness profile sampled by the $\theta$ posteriors. We think of $\varrho_{\rm eff}$ as a ``distilled" size metric, since it reduces a complicated (and intrinsically uncertain) brightness profile into a straightforward, intuitive benchmark value.

\subsection{Conversion to Physical Parameters} \label{sec:convert}

To examine the relationship between size and luminosity, we need to work with {\it physical} (rather than {\it observational}) parameters (e.g., radii in AU rather than arcseconds, and luminosities rather than flux densities) that account for both systematics in the flux calibration and the different target distances in the sample.  Our procedure for the former is to multiply each posterior sample of $F_\nu$ by a factor $s$, drawn from a normal distribution $\sim$$\mathcal{N}(1.0, 0.1)$, that mimics the $\sim$10\%\ uncertainty in the absolute flux calibration of the data.  For the latter, we start with a parallax measurement for each target and, presuming a normal probability distribution, $\sim$$\mathcal{N}(\varpi, \sigma_\varpi)$, we adopt the formalism of \citet{astraatmadja16} to infer a distance posterior, $p(d | \varpi, \sigma_\varpi)$, for a uniform distance prior.  

\begin{figure}[t!]
\includegraphics[width=\linewidth]{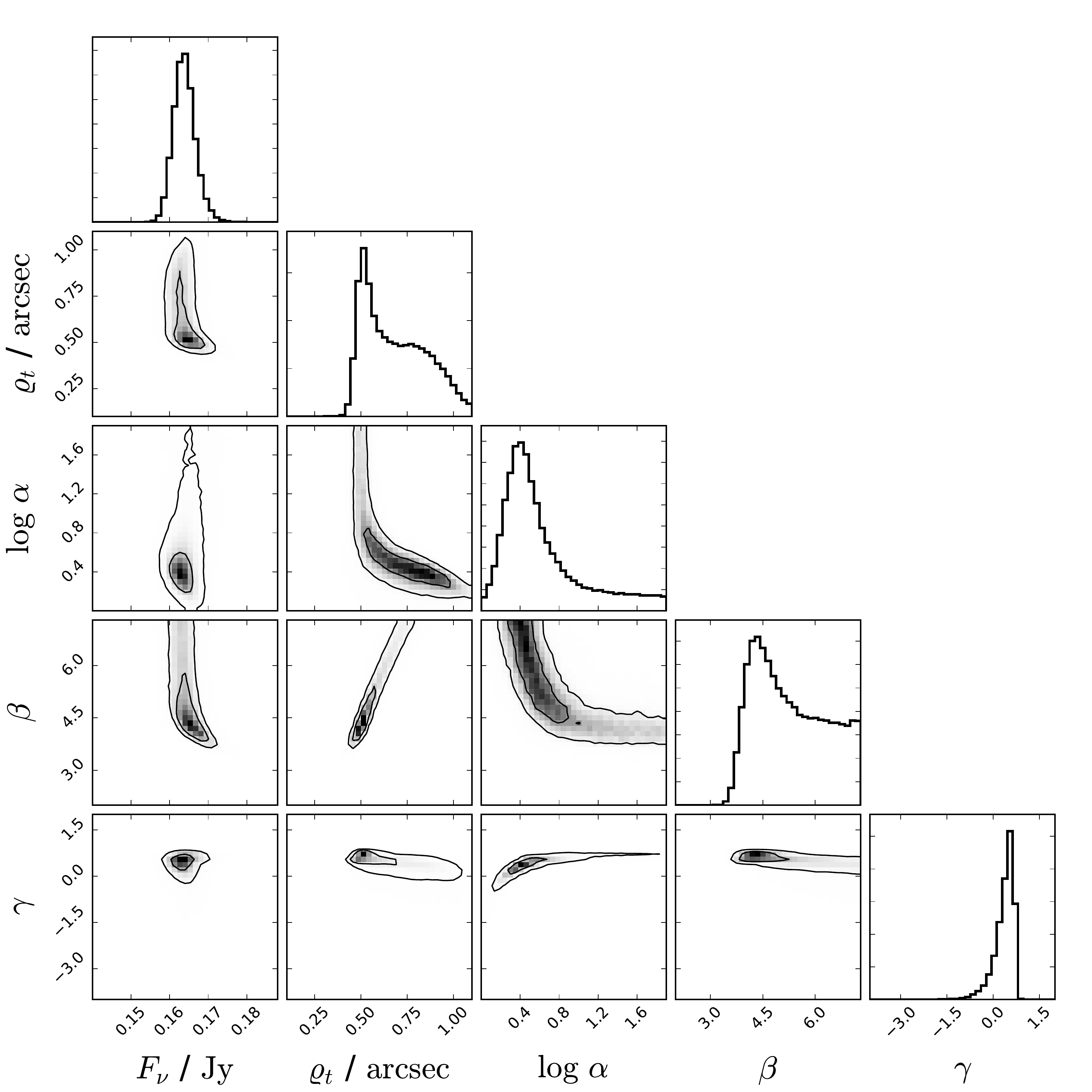}
\figcaption{Covariances of Nuker profile parameters for the IQ Tau disk.  Contours mark the 68 and 95\%\ confidence intervals.  
\label{fig:iqtau_covar}}
\end{figure}

Trigonometric parallaxes from the revised {\it Hipparcos} \citep{vanleeuwen07} and {\it Gaia} DR1 \citep{gaia_dr1} catalogs are available for AS 209, CQ Tau, UX Tau A, SU Aur, HD 163296, IM Lup, MWC 758, SAO 206462, and TW Hya.  We assign the same parallax for HP Tau that was measured for its wide companion, HP Tau/G2, using VLBI radio observations \citep{torres09}.  The remaining 22 targets in Taurus are assigned parallaxes based on the values of their nearest neighbors.  To do that, we compiled a list of 38 Taurus members with either {\it Gaia} DR1 or VLBI \citep{loinard07,torres09} trigonometric parallaxes (A.~L.~Kraus, {\it private communication}).  For each target, we calculated the mean and standard deviation of the parallaxes from those members within a 5\degr\ radius.  We associate WaOph 6 with its neighbor AS 209, albeit with an inflated $\sigma_\varpi$.  For the remaining 14 targets in Ophiuchus, we adopt the mean $\varpi$ measured from VLBI data for L1688 or L1689 sources by \citet{ortiz-leon16}.  The parallax uncertainty for DoAr 33 was increased, since it lies north of L1688.  LkH$\alpha$ 330, RX J1604.3$-$2130, and RX J1615.3$-$3255 are assigned parallaxes based on their affiliated cluster means \citep[Per OB2, Upper Sco, and Lupus, respectively; see][]{dezeeuw99,galli13}.

\begin{figure}[t!]
\includegraphics[width=\linewidth]{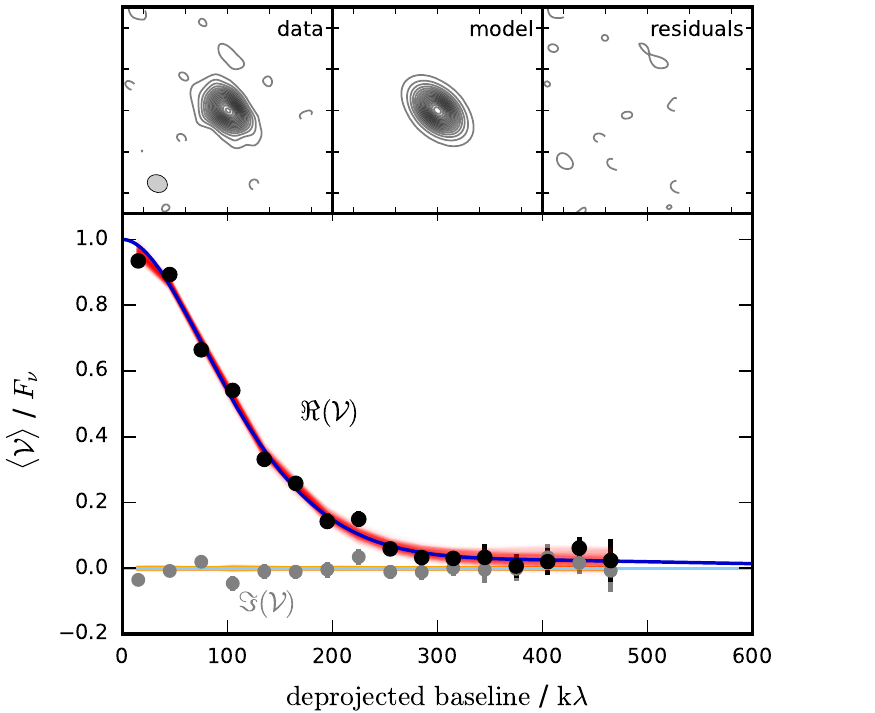}
\figcaption{(top) Synthesized images of the IQ Tau data (left), a representative model (center), and the residuals (right), as in Fig.~\ref{fig:gallery}.  (bottom) The azimuthally-averaged visibilities as a function of deprojected baseline length (in black/gray for real/imaginary components), normalized to the most probable $F_\nu$ (0.163\,Jy), overlaid with the analogous profiles derived from 200 random draws from the $\theta$ posteriors (in red/orange for real/imaginary).  The mean of those random draws is also used to construct a representative profile (in blue/cyan) that is perfectly sampled in Fourier-space and extrapolated to the full range of spatial scales in the plot.     
\label{fig:iqtau_dmr}}
\end{figure}

\begin{figure}[!b]
\includegraphics[width=\linewidth]{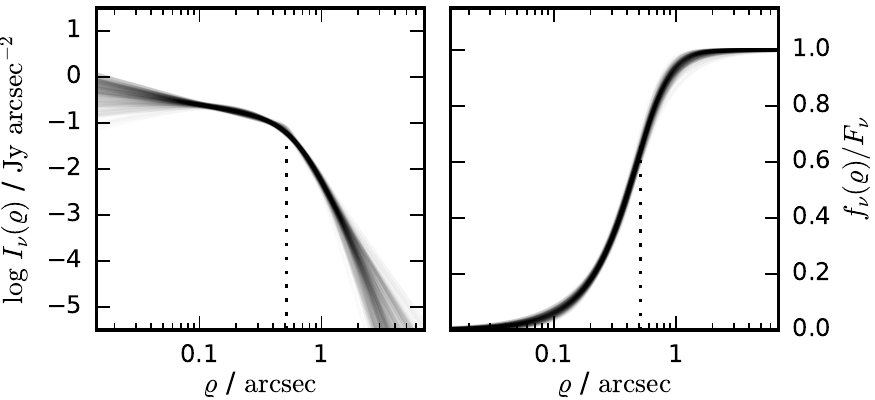}
\figcaption{The $I_\nu(\varrho)$ and $f_\nu(\varrho)/F_\nu$ profiles (as in Fig.~\ref{fig:sizemeaning}) for 200 random draws from the $\theta$ posteriors for the IQ Tau disk.  The most probable $\varrho_{\rm eff}$ is marked with a dotted vertical line.
\label{fig:iqtau_profiles}}
\end{figure}

For each posterior sample of \{$F_\nu$, $\varrho_{\rm eff}$\}, we draw a distance from $p(d | \varpi, \sigma_\varpi)$ and use it to calculate a posterior sample of the logarithms of \{$L_{\rm mm}$, $R_{\rm eff}\} = \{F_\nu \times (d/140)^2 \times s$, $\varrho_{\rm eff} \times d$\}, where $d$ is in pc units.\footnote{Note that $L_{\rm mm}$ is quantified in flux density units scaled to 140\,pc, to ease comparisons with other disk samples.}  The associated posteriors are summarized in Table~\ref{table:sizelum}.

\subsection{A Worked Example} \label{sec:example}

To illustrate more concretely the procedure outlined above, we present a step-by-step analysis of the IQ Tau disk, which has a continuum luminosity that is roughly the median of the sample  distribution.

We initialize the parameters (see Sect.~\ref{sec:SB}) using $F_\nu \sim \mathcal{U}(0.15, 0.19$\,Jy), $\varrho_t \sim \mathcal{U}(0.3, 0.6$\arcsec), $\log{\alpha} \sim \mathcal{U}(0.3, 1.3)$, $\beta \sim \mathcal{U}(2, 6)$, $\gamma \sim \mathcal{U}(-1, 1)$, $i \sim \mathcal{U}(44, 58\degr)$, $\varphi \sim \mathcal{U}(30, 45\degr)$, $d\alpha \sim \mathcal{U}(0.01, 0.11\arcsec)$, $d\delta \sim \mathcal{U}(-0.39, -0.29\arcsec)$.  We then sample the posterior with the MCMC algorithm, and reach a stationary target distribution after \dt{4,000} steps.  The walkers have acceptance fractions of 0.26--0.30; the autocorrelation times are 80--90 steps. 
Figure~\ref{fig:iqtau_covar} shows the inferred covariances and marginalized posteriors for the Nuker profile parameters.  We omit the geometric parameters for clarity, but they agree with independent estimates \citep[e.g.,][]{guilloteau14}.  Confidence intervals for each parameter are derived from the marginalized posteriors and noted in Table~\ref{table:SBpars}.  Figure~\ref{fig:iqtau_dmr} compares the simulated observations constructed with random draws from these posteriors to the data.        

\begin{figure*}[t!]
\includegraphics[width=7in]{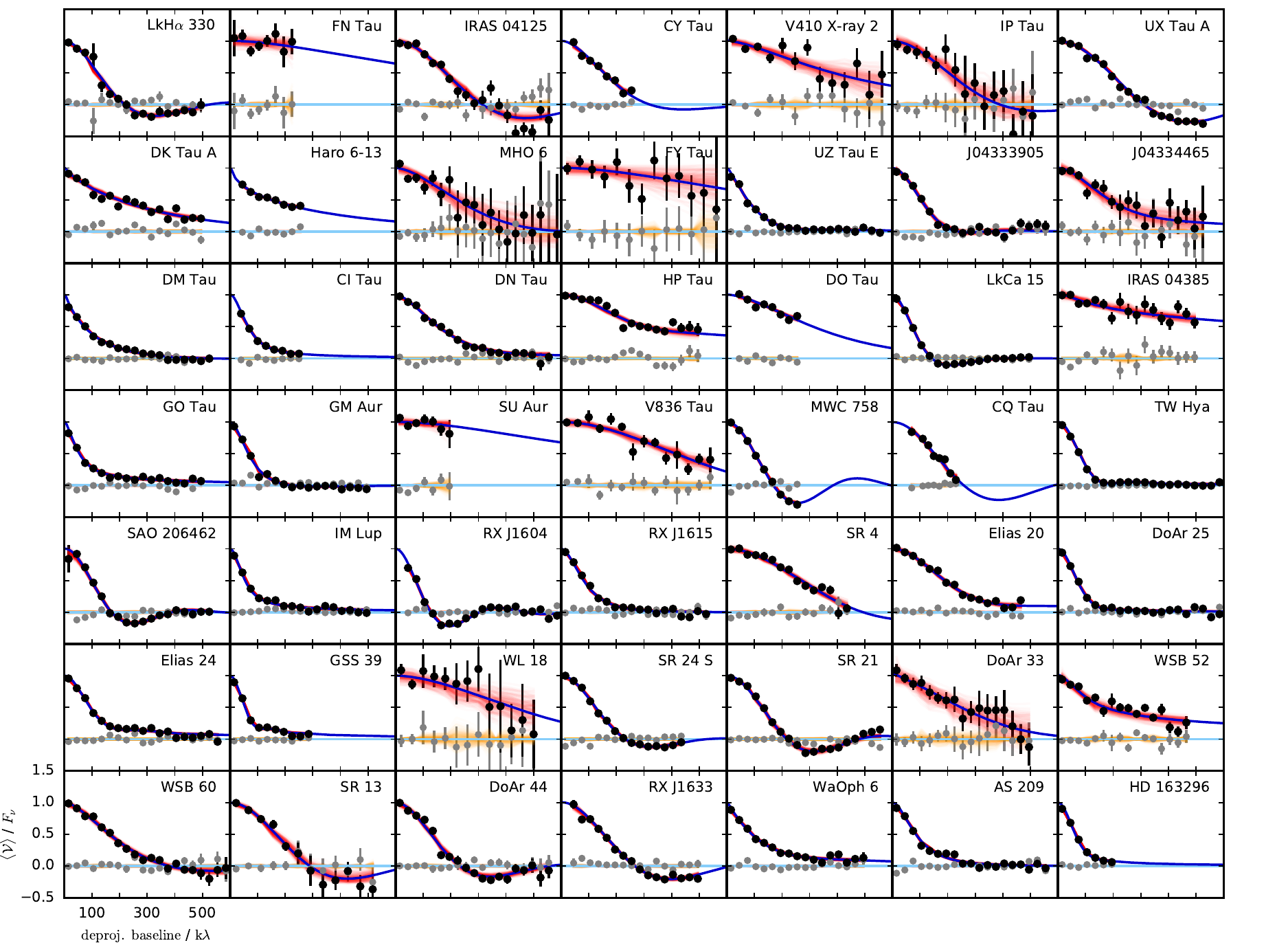}
\figcaption{The deprojected, azimuthally averaged, and normalized (by the inferred most probable values of $F_{\nu}$) real (black points) and imaginary (gray points) visibility profiles as a function of baseline length are shown as points with associated uncertainties.  The red (real) and orange (imaginary) curves show the corresponding profiles derived from 200 random draws of the posterior parameters.  The blue and cyan curves show the means of those draws, now sampled perfectly in the Fourier domain, to guide the eye.  \label{fig:vis_gallery}}
\end{figure*}

To construct posterior samples of $\varrho_{\rm eff}$, we calculate $f_\nu(\varrho)/F_\nu$ (see Eq.~\ref{eq:fcum}) for each posterior sample of $\theta$.  As an illustration, Figure~\ref{fig:iqtau_profiles} shows $I_\nu(\varrho)$ and $f_\nu(\varrho)/F_\nu$ for 200 random draws from the posteriors.    

Circling back to our arguments advocating for a size metric like $\varrho_{\rm eff}$ (see Sect.~\ref{sec:size}), rather than $\varrho_t$ (or its equivalent in other model prescriptions), IQ Tau provides a clear demonstration of how the former is more precise than the latter in the face of poorly constrained gradient parameters.  Because \{$\alpha$, $\beta$, $\gamma$\} are determined with relatively poor precision in this case, the 68\%\ confidence interval on $\varrho_t$ is \dt{(0.46, 0.90\arcsec)}.  On the contrary, the inference on $\varrho_{\rm eff}$ is considerably narrower, \dt{(0.53, 0.56\arcsec)}.  The ambiguity in the indices does not  matter much for $\varrho_{\rm eff}$: even for a diversity of $I_\nu(\varrho)$, the models that satisfactorily reproduce the data essentially have similar $f_\nu(\varrho)$.  

\dt{To emphasize the point that the functional form of the brightness profile does not matter, we compared these Nuker profile inferences to the more standard broken power-law and similarity solution models.  For the broken power-law prescription, the radius marking the transition between the two gradients lies in the range (0.40, 0.52\arcsec); for the similarity solution, the radius marking the transition between the inner power-law and the outer exponential taper is constrained to (0.22, 0.38\arcsec).  Despite these differences, all three models find {\it statistically indistinguishable} constraints on the effective size metric ($\varrho_{\rm eff}$), with a 68\%\ confidence interval (0.53, 0.56\arcsec).}

\section{Results} \label{sec:results}

\begin{figure*}
\includegraphics[width=7in]{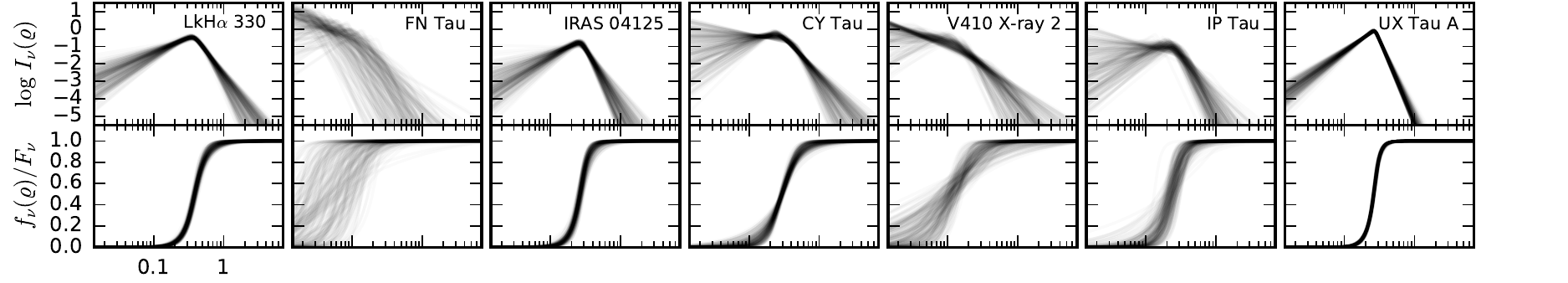}
\includegraphics[width=7in]{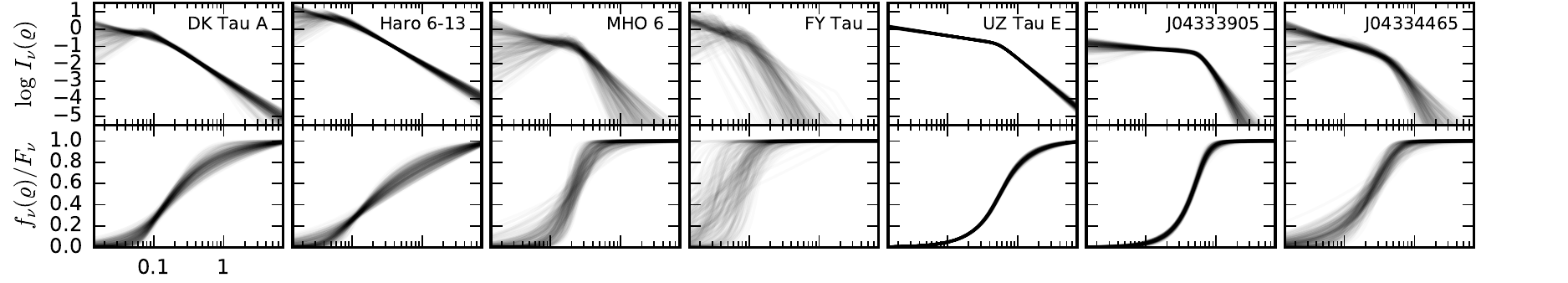}
\includegraphics[width=7in]{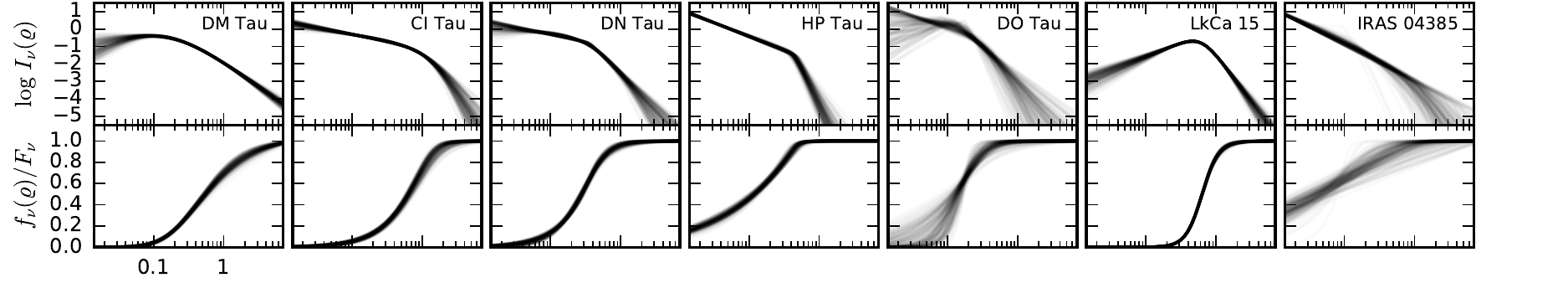}
\includegraphics[width=7in]{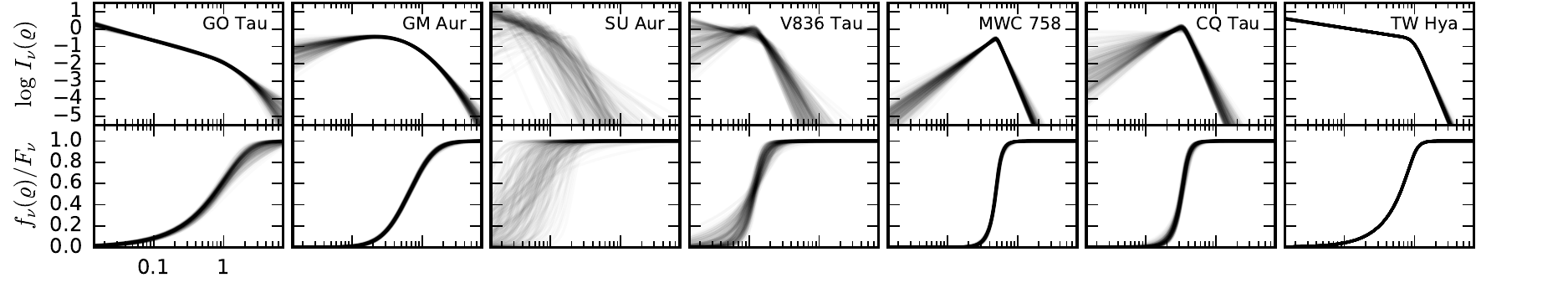}
\includegraphics[width=7in]{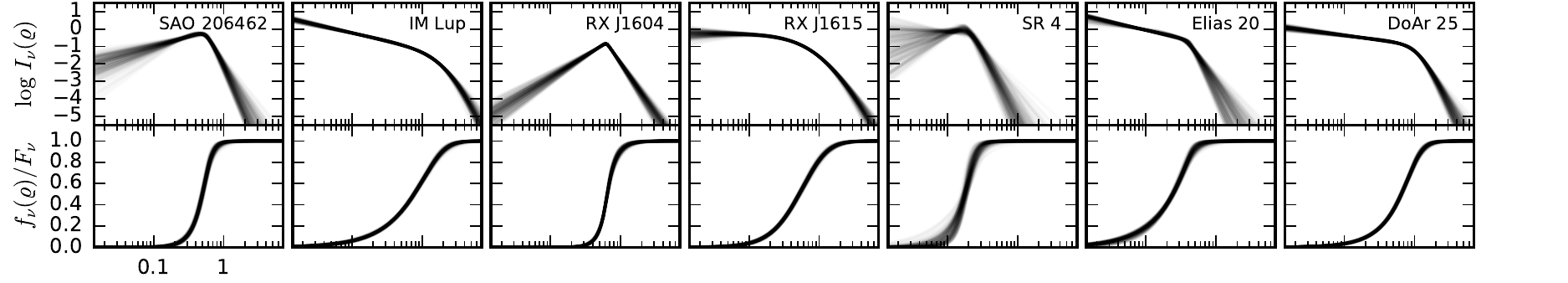}
\includegraphics[width=7in]{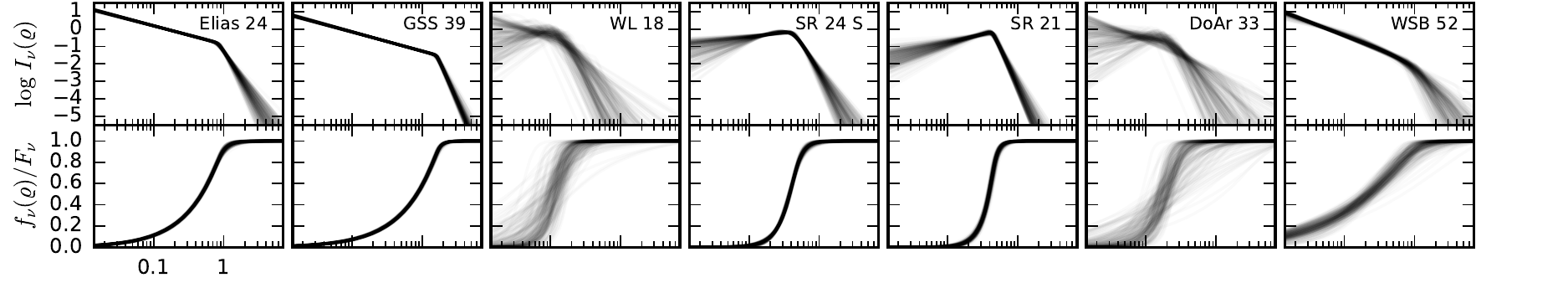}
\includegraphics[width=7in]{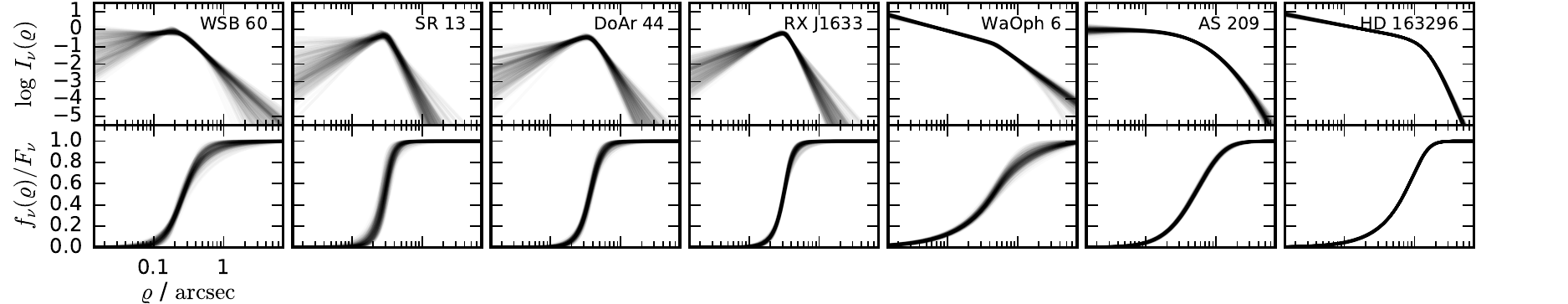}
\figcaption{The $I_{\nu}(\varrho)$ profiles (top panels of each row; in Jy arcsec$^{-2}$ units) and $f_{\nu}(\varrho)/F_{\nu}$ profiles (bottom panels of each row) constructed from 200 random draws from the posteriors for each disk target (cf., Fig.~\ref{fig:iqtau_profiles}).   \label{fig:profiles}}
\end{figure*}

Having demonstrated the modeling procedure with a representative example target, we now present the results for the rest of the sample.  A condensed summary of the inferred posterior distributions for the surface brightness model parameters and $\varrho_{\rm eff}$ is provided in Table~\ref{table:SBpars}.  Figure~\ref{fig:vis_gallery} makes direct comparisons between the posteriors and the observed SMA visibilities.  The surface brightness profiles, $I_{\nu}(\varrho$), and cumulative intensity profiles, $f_\nu(\varrho)/F_\nu$, are shown together in Figure~\ref{fig:profiles}. 

\begin{figure}
    \includegraphics[width=\linewidth]{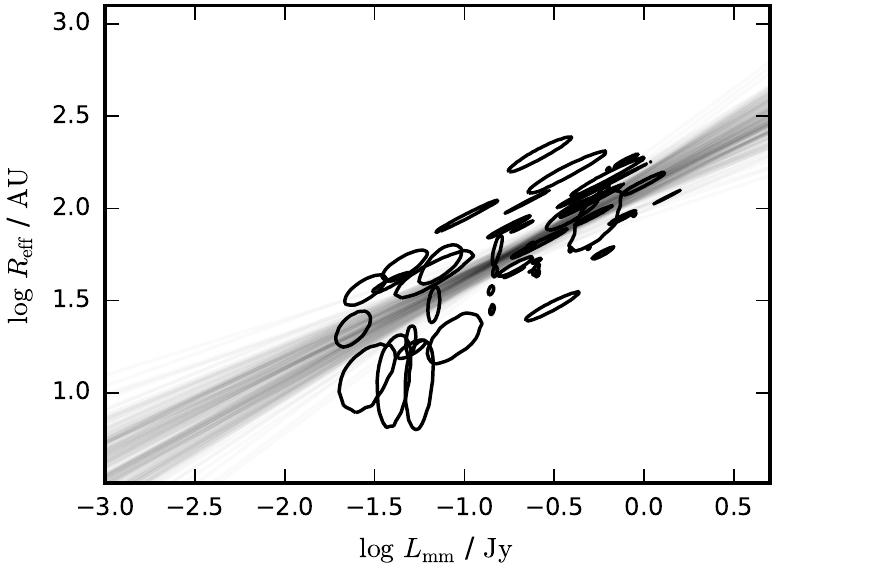}
    \figcaption{Disk size (as defined in Sect.~\ref{sec:size}) as a function of the 340\,GHz luminosity.  Each ellipse represents the 68\%\ joint confidence interval on $\{\log{R_{\rm eff}}, \log{L_{\rm mm}}\}$.  The gray curves show 200 random draws from the linear regression posteriors.  
    \label{fig:lumsize}}
\end{figure}

Figure \ref{fig:lumsize} shows the resulting size--luminosity relation.  As was hinted at in previous studies \citep{andrews10b,pietu14}, we find a significant correlation between these {\it empirical} variables that coarsely describe the continuum intensity profile, such that brighter disks have their emission distributed to larger radii.  We employed the \citet{kelly07} linear regression mixture model to quantify the relationship and found that
\begin{equation}
\dt{\log{R_{\rm eff}} = (2.12\pm0.05) + (0.50\pm0.07) \log{L_{\rm mm}}},
\label{eq:sizelum}
\end{equation}
with a Gaussian scatter perpendicular to that scaling with a standard deviation of $0.19\pm0.02$\,dex (where $R_{\rm eff}$ is in AU and $L_{\rm mm}$ is in Jy at an adopted distance of 140\,pc; all uncertainties are quoted at the 68\%\ confidence level).  The strength of this correlation, its slope, and the amount of scatter around it are essentially the same for a considerable range of $R_{\rm eff}$ definitions (e.g., specifically for $x \in [0.5, 0.8]$, see Sect.~\ref{sec:size} for details\footnote{As one would expect, the intercept values depend on the adopted $x$.  Technically the relationship inferred here is consistent with the data regardless of the choice of $x$, although for values of $x$ much outside the quoted range, the uncertainties on $R_{\rm eff}$ grow large enough that the constraints on the slope are rather poor.}).   

Perhaps the most surprising aspect of this relationship is the slope, which indicates that $R_{\rm eff} \propto \sqrt{L_{\rm mm}}$.  That behavior has two striking (and related) implications: (1) the luminosity scales with the emitting area and (2) the surface brightness averaged over the area inside $R_{\rm eff}$ is roughly constant for all luminosities.  With respect to the latter point, we can derive 
\begin{eqnarray}
\langle I_{\nu} \rangle|_{<R_{\rm eff}} &=& \frac{\int_0^{\rm \varrho_{\rm eff}} I_{\nu}(\varrho) \, 2\pi \, \varrho \, d\varrho}{\int_0^{\rm \varrho_{\rm eff}} \, 2\pi \, \varrho \, d\varrho} = \frac{x F_\nu}{\pi \varrho_{\rm eff}^2}  \nonumber \\
&=& \frac{x L_{\rm mm}}{\pi (R_{\rm eff}/140)^2} \approx \frac{x 10^{-4.3} R_{\rm eff}^2}{\pi (R_{\rm eff}/140)^2} \label{eq:meanderiv}
\\
&\approx& 0.2 \, {\rm Jy} \, {\rm arcsec}^{-2}. \nonumber
\end{eqnarray}
The first line of Eq.~(\ref{eq:meanderiv}) defines a surface brightness average and employs Eq.~(\ref{eq:fcum}); the second line uses the definitions of $L_{\rm mm}$ (Jy) and $R_{\rm eff}$ (AU; see Sect.~\ref{sec:convert}) and folds in the approximate relationship inferred in Eq.~(\ref{eq:sizelum}) (assuming the logarithmic slope is exactly 1/2); and the third line substitutes in the relevant numerical values.  That average surface brightness corresponds to an average brightness temperature of $\langle T_b \rangle|_{<R_{\rm eff}} \approx 8$\,K.

The same behavior is apparent regardless of the inferred emission morphologies.  Most strikingly, the ``transition" disks \dt{(around LkH$\alpha$ 330, IRAS 04125+2902, UX Tau A, DM Tau, LkCa 15, GM Aur, MWC 758, CQ Tau, TW Hya, SAO 206462, RX J1604.3-2130, RX J1615.3-3255, SR 24 S, SR 21, WSB 60, DoAr 44, and RX J1633.9-2422)} follow the same size--luminosity trend as the ``normal" disks.  To better quantify that finding, we compared the effective radii with a concentration parameter, defined as the ratio of effective radii for different $x$ (e.g., the ratio of the radii that encircle the first and third quantiles of the total luminosity).  But as one might expect, we find no obvious connections between $R_{\rm eff}$ or $L_{\rm mm}$ and such a concentration metric.\footnote{Although it should be obvious that the ``transition" disks have comparatively high concentrations at a given $R_{\rm eff}$.}

\section{Discussion} \label{sec:discussion}  

We have compiled a resolved survey of the 340\,GHz (880\,$\mu$m) continuum emission from 50 nearby protoplanetary disks using new and archival SMA observations.  Our primary goal was to validate and better characterize the putative trend between continuum sizes and luminosities first identified by \citet{andrews10b}, both by expanding the survey size (by a factor of 3) and extending down to intrinsically less luminous disk targets (by a factor of 2).  To interpret these data in this context, we modeled the observed visibilities with a simple, flexible surface brightness prescription and used the results to establish a standardized ``size" metric.  These sizes, based on the radius ($R_{\rm eff}$) that encircles a fixed fraction of the continuum luminosity ($L_{\rm mm}$), are intuitive, empirical quantities that are robust to any intrinsic uncertainties in the slope(s) of the surface brightness profiles. 

We find a significant correlation between $L_{\rm mm}$ and $R_{\rm eff}$, such that the luminosity scales with the effective emitting area.  That size--luminosity relation suggests that disks have a roughly constant average 340\,GHz surface brightness inside their effective radii, $\sim$0.2\,Jy\,arcsec$^{-2}$ (8\,K in terms of a brightness temperature). 

The shape of that scaling is approximately the same as was identified by \citet{andrews10b} using a subset of the same data as in this sample, albeit in that case for the peripherally related quantities of disk mass and ``characteristic" size.\footnote{In some sense, those radii are more similar to $\varrho_t$ than $\varrho_{\rm eff}$, and thereby subject to the same issues identified in Sect.~\ref{sec:size}; see Fig.~\ref{fig:sizemeaning}.}  \citet{pietu14} also claimed a size--luminosity trend in their analogous 230\,GHz (1.3\,mm) sample \citep[$\sim$25 disks, including data from][]{guilloteau11}, although they did not quantify it.  Using their flux densities and characteristic sizes for a regression analysis as conducted in Section~\ref{sec:results}, we confirm a marginal ($\sim$3\,$\sigma$) correlation that indeed also has the same slope as inferred for the SMA 340\,GHz sample.

It is also interesting to note that $\sim$0.2\,dex of (presumed to be Gaussian) dispersion perpendicular to this scaling is required to reproduce the observed scatter of the individual measurements beyond their inferred uncertainties.  The regression analysis used to infer that scatter does not provide information on its underlying cause.  However, one compelling possibility is that it may be related to the diversity of heating the disks might experience given the wide range of stellar host (irradiation source) properties included in this sample.

This stellar heating hypothesis is testable, at least in a crude sense.  We can do that by re-scaling the $L_{\rm mm}$ inferences by a factor $\propto 1/B_\nu(\langle T_d \rangle)$, where $B_\nu$ is the Planck function and $\langle T_d \rangle$ is a rough estimate of the average dust temperature inside the effective radius.  Ideally, we could calculate $\langle T_d \rangle$ weighted by the continuum optical depth \citep[e.g.,][]{andrews13}, but since we have opted for an empirical modeling approach here, such information is not directly accessible.  As an approximation, we can presume the emission is optically thin and thereby adopt a weighting function $w(r) = I_{\nu}(r)/B_\nu[T_d(r)]$, so that
\begin{equation}
\langle T_d \rangle = \frac{\int_0^{R_{\rm eff}} w(r) \, T_d(r) \, 2\pi \, r \, dr}{\int_0^{R_{\rm eff}} w(r) \, 2\pi \, r \, dr}.
\label{eq:meantd}
\end{equation}
Now we need to specify a presumed parametric form for $T_d(r)$ that roughly captures the behavior of irradiation heating by the central star.  We base that behavior on the analysis of a suite of radiative transfer models of representative disks by \citet{andrews13} and set
\begin{equation}
T_d(r) \approx T_{10} \left(\frac{L_\ast}{L_\odot}\right)^{1/4} \left(\frac{r}{10 \, {\rm AU}}\right)^{-1/2}
\label{eq:tsimple}
\end{equation}
where $T_{10} = 30\pm5$\,K is the temperature at 10\,AU and $L_\ast$ is the stellar host luminosity.  We also impose a floor on $T_d(r)$, such that it cannot go below a background level of 7\,K (the adopted value makes little difference, so long as it is in a reasonable range, $\lesssim 12$\,K).  Using this prescription, we collated $L_\ast$ estimates from the literature\footnote{The adopted $L_\ast$ values were appropriately scaled to the draws from the distance posteriors.} \citep[see][]{testi03,natta04,andrews10b,andrews11,andrews12,andrews13,cieza12,barenfeld16,cleeves16} and computed posterior samples of $\langle T_d \rangle$ for each posterior draw of brightness profile parameters, including a representative 20\%\ uncertainty in $L_\ast$ (along with the distance and $T_{10}$ uncertainties).  Table~\ref{table:sizelum} includes the derived $\langle T_d \rangle$ and adopted $L_\ast$ values.  

With this coarse metric of the disk heating in hand, we can then examine the relationship between $R_{\rm eff}$ and a quantity $\propto L_{\rm mm}/B_{\nu}(\langle T_d \rangle)$ that should account for any scatter introduced by dust heating in the analysis of the size--luminosity relation.  For the sake of a familiar comparison, we opt to re-cast the scaled luminosity dimension in the context of a disk mass estimate.  Again assuming the emission is optically thin, we followed \citet{andrews13} to calculate 
\begin{equation}
\log{M_d} = \log{\left[\frac{L_{\rm mm}}{B_{\nu}(\langle T_d \rangle)}\right]} + 2\log{d^{\prime}} - \log{(\zeta \kappa_\nu)},
\end{equation}
where $d^{\prime} = 140$\,pc (the adopted reference distance for $L_{\rm mm}$), $\zeta = 0.01$ (an assumed dust-to-gas mass ratio), and $\kappa_\nu = 3.5$\,cm$^2$\,g$^{-1}$ (a standard 340\,GHz dust opacity).

\begin{figure}[t!]
    \includegraphics[width=\linewidth]{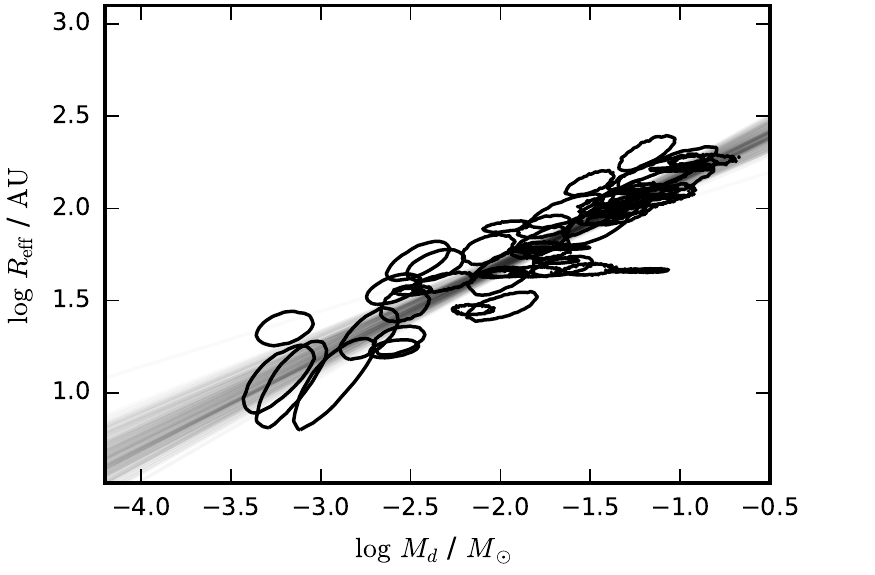}
    \figcaption{Disk size (as defined in Sect.~\ref{sec:size}) as a function of the total disk mass (see text for associated assumptions).  Each ellipse represents the 68\%\ joint confidence interval on $\{\log{R_{\rm eff}}, \log{M_d}\}$.  The gray curves show 200 random draws from the linear regression posteriors.  There is notably less scatter than in its scaling with $L_{\rm mm}$ (see Fig.~\ref{fig:lumsize}), suggesting that much of the dispersion in that relation can be explained by  a heating (temperature) effect.  \label{fig:sizemass}}
\end{figure}

Figure \ref{fig:sizemass} shows the corresponding relationship between the effective radii and these masses.  It is immediately apparent that the correlation is stronger and tighter after accounting for $\langle T_d \rangle$, at least partially because much of the perpendicular spread has been redistributed {\it along} the trend.  We performed the same regression analysis in the $\{\log{R_{\rm eff}}, \log{M_d}\}$-plane as before and found
\begin{equation}
\dt{\log{R_{\rm eff}} = (2.64\pm0.07) + (0.47\pm0.04)\times \log{M_d}}
\end{equation}
(where again $R_{\rm eff}$ is in AU and $M_d$ is in $M_\odot$ units; the quoted uncertainties are at the 68\%\ confidence level).  The scatter perpendicular to this relation is considerably smaller than in Figure~\ref{fig:lumsize}: it has a (presumed Gaussian) dispersion of $\sim$0.10\,dex, but is consistent with the scatter arising solely from the data uncertainties (at 2.5$\sigma$).  

This demonstrates that the inferred {\it slope} of the size--luminosity relation is preserved (and indeed reinforced) by the temperature correction, which in this framework implies a roughly constant average surface density interior to $R_{\rm eff}$, $\langle \Sigma \rangle|_{< R_{\rm eff}} \approx 10$\,g\,cm$^{-2}$ (following the same logic as in Eq.~\ref{eq:meanderiv}).  That value can be re-cast into an average optical depth, $\langle \tau \rangle|_{< R_{\rm eff}} \approx 1/3$.  

We will revisit that optical depth inference below, but it is worth pointing out that these results do not depend much on the choice for the weighting function in Eq.~(\ref{eq:meantd}).  If we instead adopt an empirical $w(r) = I_{\nu}(r)$ or something more appropriate for optically thick emission, $w(r) = B_{\nu}(T_d)$, the inferred slope and scatter of the relationship are not significantly different.\footnote{Naturally the intercepts change to reflect the shifts in $\langle T_d \rangle$.}

\subsection{Potential Origins of a Size--Luminosity Relation}

While the existence and qualities of a mm continuum size--luminosity relationship are now emerging, it is not particularly obvious {\it why} we might expect such behavior.  The preliminary speculations on its potential origins can be differentiated into two broad categories: (1) the scaling is a natural consequence of the initial conditions, potentially coupled with some dispersion introduced by evolutionary effects; or (2) the scaling reflects some universal structural configuration.  The concepts behind these categories are not always distinct, and certainly not mutually exclusive, but in light of the new data analysis presented here we will explore them separately in the following sections.

\subsubsection{Initial Conditions, Evolutionary Dispersion}

An appeal to a specific distribution of initial conditions may seem like a fine-tuning solution, but it cannot be easily dismissed given the strong links expected between young disks and the star formation process.  \citet{andrews10b} noted that the size--luminosity relationship is nearly perpendicular to the direction expected for viscously evolving disks that conserve angular momentum, and because of that they speculated that it may point to the underlying spread in the specific angular momenta in molecular cloud cores \citep[see also][]{isella09}.  A few other possibilities along these lines were raised by \citet{pietu14}, including tidal truncation (unlikely for this sample of primarily single stars in low-density clusters; but see \citealt{testi14}) and significantly broad age and/or viscosity distributions.  But neither of these studies considered how those initial conditions in the {\it gas} disks would be manifested a few million years later in observational tracers of the solid particles.   

\begin{figure}[t!]
    \includegraphics[width=\linewidth]{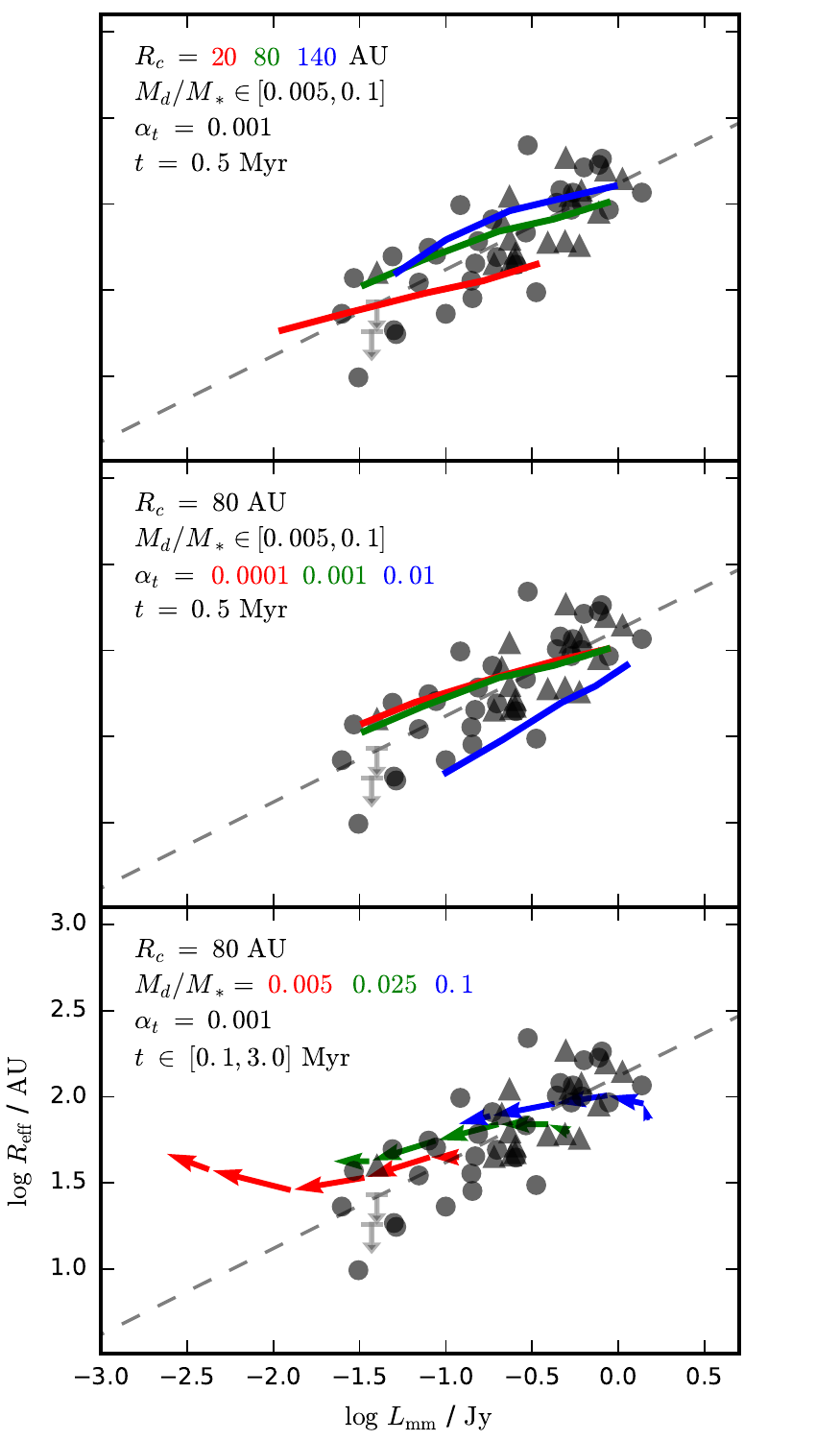}
    \figcaption{Comparisons of the measured size--luminosity relationship with simple models for the evolution of disk solids with varying initial conditions.  In all panels, the data are shown as gray points \dt{(``normal" disks) or triangles (``transition" disks)} at the peaks of the joint posterior distributions (or arrows for upper limits).  The dashed line shows the best-fit regression inferred in Sect.~\ref{sec:results}.  The colored curves show 3 different sets of demonstrative models that produce a comparable trend: ({\it top}) a snapshot at 0.5\,Myr for a range of disk masses, a fixed viscosity coefficient, and three  characteristic radii; ({\it middle}) as in the top, but now for a single fixed $R_c$ and three viscosity coefficients; ({\it bottom}) the time evolution for disks with three  masses -- the tip of each arrow represents a time step in the series [0.2, 0.5, 1, 2, 3]\,Myr.    
    \label{fig:models}}
\end{figure}

To assess this category of explanation for the size--luminosity relation, we considered the expected behavior of disks in the \{$L_{\rm mm}$, $R_{\rm eff}$\}-plane for simple models of the evolution of their solids when embedded in a standard viscously evolving gas disk.  We adopted the methodology of \citet{birnstiel15} to roughly estimate how the radial surface density profiles for solids of different sizes evolve with time for a range of initial conditions, parameterized through the total disk masses ($M_d$), (initial) characteristic radii ($R_c$), and (fixed) viscosity coefficients ($\alpha_t$).  In these calculations, we assumed a representative 0.5\,$M_{\odot}$ host star and a fragmentation velocity of 10\,m\,s$^{-1}$.  For each time and combination of parameters from a coarse grid, we constructed a radial profile of 340\,GHz optical depths as the summed (over particle size) product of the computed surface densities and representative opacities (the latter using the prescription of \citealt{ricci10a}).  We then converted that to an intensity profile with a crude approximation, $I_\nu (r) = B_\nu [T(r)] (1 - e^{-\tau_\nu(r)})$, using a (fixed) representative temperature profile appropriate for the presumed stellar host (see Eq.~\ref{eq:tsimple}).  The luminosities and effective radii corresponding to those intensity profiles were calculated as in Sections~\ref{sec:size} and \ref{sec:convert}.     

Figure~\ref{fig:models} shows a highly condensed summary of the models compared with the size and luminosity metrics inferred from the data.  These models actually do a reasonably good job at reproducing the observed mean trend with a $M_d/M_\ast$ distribution from $\sim$0.5--10\%, a modest range of initial $R_c$ of $\sim$60--150\,AU, and relatively low turbulence levels, $\alpha_t \lesssim 0.001$, but preferentially at {\it early} points in the evolutionary sequence ($<$1\,Myr).    

While individual young star ages are notoriously uncertain \citep[e.g.,][]{soderblom14}, the nominal cluster ages from which this sample is drawn are thought to be more like 2--3\,Myr.  By those times, the model predictions have diverged up and to the left in the \{$L_{\rm mm}$, $R_{\rm eff}$\}-plane \dt{(away from the mean relation, with larger $R_{\rm eff}$ than $>$90\%\ of the sample targets for a given $L_{\rm mm}$)} due to both viscous spreading and a relative depletion of particles that emit efficiently in the mm continuum.  That timescale discrepancy is not surprising; it is another manifestation of a generic feature of such models that assume a smooth gas disk structure \citep[e.g.,][]{takeuchi02,takeuchi05,brauer07}.

Putting aside the timescale problem, it is interesting that these models tend to predict a shallower slope than what is inferred from the data.  This might suggest a correlation between the initial $M_d$ and $R_c$, perhaps as a signature of the disk formation process.  Alternatively, it might indicate a disk mass-dependent evolutionary timescale (perhaps a relationship between $M_d$ and $\alpha_t$).  Unfortunately, it would be difficult to disentangle such effects from the stellar age and mass biases likely present in such an inhomogeneously-selected sample.

\subsubsection{Small-Scale, Optically Thick Substructures}

The timescale discrepancy highlighted above may be a telling failure of the underlying assumptions used to set up such models.  The remedy often proposed for this inconsistency is the presence of fine-scale, localized maxima in the radial pressure profile of the gas disk \citep[e.g.,][]{whipple72,pinilla12a}.  Those pressure peaks attract drifting particles, can slow or stop their inward motion, and thereby produce high solid concentrations in small areas that may promote rapid growth to larger bodies.  Those concentrations would likely be manifested in the mm continuum as small regions of optically thick emission.  Indeed, recent studies have found that narrow rings of bright emission accompanied by darker gaps, among other features, are prevalent in the few disks that have already been observed at very high spatial resolution \citep{brogan15,zhang16,andrews16,cieza16,perez16,isella16,loomis17}.  Such fine-scale optically thick features were considered by \citet{pietu14} as a potential contributor to a size--luminosity relation.  Having firmly established the character of that relation here, it makes sense to revisit that possibility.

\begin{figure}[t!]
    \includegraphics[width=\linewidth]{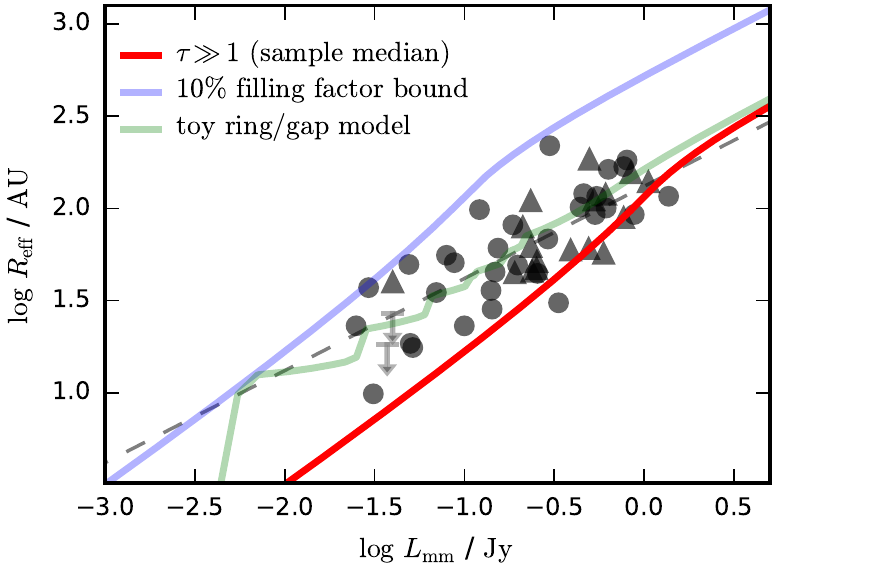}
    \figcaption{Comparisons of the measured size--luminosity relation with simple models for the optically thick disk structures; symbols are as in Fig.~\ref{fig:models}.  The red curve is a purely optically thick model with the sample median temperature profile as given in Eq.~(\ref{eq:tsimple}).  The blue curve corresponds to that same model with a (uniformly-distributed) 10\%\ (areal) filling factor.  The inferred distribution of disks in the \{$L_{\rm mm}$, $R_{\rm eff}$\}-plane could be reproduced by optically thick regions with filling factors of a few tens of percent.  The green curve is a toy variant of the pure optically thick model, with annular segments of the disk removed to mimic substructures like those recently seen in very high resolution images.  \label{fig:thick}}
\end{figure}

In Figure~\ref{fig:thick} we illustrate how optically thick emission could mimic the inferred distribution of disks in the \{$L_{\rm mm}$, $R_{\rm eff}$\}-plane.  A fiducial optically thick surface brightness profile was constructed by setting $I_\nu(r) = B_\nu[T_d(r)]$, where $T_d(r)$ is specified from the approximation in Eq.~(\ref{eq:tsimple}).  We then consider a suite of such models where the disk extends out to some sharp cutoff radius, and for each of those profiles calculate a luminosity and effective size as described in Section~\ref{sec:model}.  For the sample median stellar luminosity ($\sim$1\,$L_{\odot}$), this results in the red curve in Figure~\ref{fig:thick}.  In this context, it makes sense that we do not see disks much to the right of that boundary (since they would all be completely optically thick and therefore cannot produce more emission than allowed by the Planck function).  If a given disk has an optically thin contribution or small-scale substructures that reduce the filling factor of optically thick emission below unity, it will lie to the left of this fiducial curve (as illustrated by the blue curve for the case of a 10\%\ filling factor).     

In this scenario, the inferred distribution of disks in the size--luminosity plane could be explained if \dt{disks have} optically thick filling fractions (inside a given $R_{\rm eff}$) of a few tens of percent.  As one might expect, that is naturally in line with the rough estimate of the mean optical depth ($\approx$1/3) inside $R_{\rm eff}$ implied by the inferred size--luminosity relation earlier in this section.  \dt{Adopting again the IQ Tau disk as an example, we find that its location in the size--luminosity plane could be explained with an optically thick filling factor of $\sim$10--30\%\ inside 80\,AU.}

Of course, the shift away from the pure optically thick curve for any  disk can be achieved in myriad ways, depending on the detailed spatial distribution of the emission.  That said, it is instructive to note that the mean size--luminosity relation can, in principle, be reproduced with a toy model of rings and gaps, as shown by the green curve in Figure~\ref{fig:thick}.  There, we have intentionally tuned the model to match the inferred relation by simply removing regions of emission from the purely thick model that span $\sim$few AU segments at regular spacings.  The intent is to crudely mimic recent high resolution images of disks.  In this case, brighter disks represent systems where the local pressure maxima continue to larger distances from the host star (i.e., brighter disks are just systems that have more large rings).  It is worth highlighting that one of the key hints that, for us, makes this explanation more than just plausible is that both ``normal" and ``transition" disks populate the same size--luminosity trend.  The latter systems could be considered an especially simple form of substructure.         

A robust assessment of the viability of this explanation will require measurements with much higher angular resolution than are yet available.  But if such data end up demonstrating that optically thick substructures are important contributors to the continuum emission, the size--luminosity relation could end up being a useful (albeit indirect) demographic metric for understanding their origins in the broader disk population (especially because it is far less observationally expensive to reconstruct the \{$L_{\rm mm}$, $R_{\rm eff}$\}-plane than to measure fine-scale emission distributions for large samples).

If this explanation is relevant, it creates a series of other peripheral issues.  For example, if optical depths are high, it means previous assessments of the disk masses might instead just reflect the underlying spatial distribution of substructures.  It could imply that the inferences of particle growth from continuum spectral indices are contaminated; indeed, \citet{ricci12} already warned of essentially this same scenario.  And it might indicate that the interpretation of optically thin spectral line emission from key tracer molecules near the inner disk midplane could be complicated (it could be obscured by the optically thick continuum, depending on its vertical distribution; e.g., \citealt{oberg15,cleeves16}).

\subsection{Caveats and Future Work}

One concern that limits our conclusions stems from the inhomogeneous sample of targets available in the SMA archive.  This sample is notable for its size, but is composed of disks \dt{in the bright half of the mm luminosity distribution}, which belong to various clusters (and thereby potentially different ages/environments) and consider a limited range of host masses.  It would be beneficial to repeat this analysis for targets \dt{with lower continuum luminosities and host masses, and that} formed in the same environment at the same time (or nearly so).   An investigation for an {\it older} cluster (say $\sim$5\,Myr) would be particularly compelling.  The evolutionary models described above predict that the size--luminosity relationship should flatten out: older disks have more radially concentrated solids, so smaller particles that tend to emit less overall in the mm continuum, but over more extended distributions, end up dominating the behavior in the \{$L_{\rm mm}$, $R_{\rm eff}$\}-plane.  In the substructure scenario, we might expect little change in the shape of the relation, although effects like growth in the trapped regions could make it difficult to characterize any trend.    

Another important point to keep in mind is that the relationship we have quantified here is expressly only valid at $\sim$340\,GHz.  While \citet{pietu14} appear to have tentatively found a similar correlation at $\sim$230\,GHz, we know empirically that the spatial extents of continuum emission from disks tend to increase with the observing frequency \citep[e.g.,][]{perez12,lperez15b,menu14,tazzari16}.  The frequency-dependent variation in the size--luminosity relation that this behavior might imply could prove to be crucial in better understanding its origins.  In addition to improving the sample quality, complementing the analysis with analogous measurements at a lower ($\lesssim$100\,GHz) frequency (albeit at higher angular resolution, given the observed tendency for more compact emission at lower frequencies) should also be viewed as highly desirable.

\acknowledgments We thank John Carpenter, Antonella Natta, and an anonymous reviewer for their helpful comments on drafts of this article, as well as Ian Czekala, Mark Gurwell, and Adam Kraus for providing useful advice.  This research greatly benefited from the {\tt Astropy} \citep{astropy} and {\tt Matplotlib} \citep{matplotlib} software suites, the {\tt emcee} package \citep{foreman-mackey13} and {\tt corner} module \citep{corner}.  We are grateful for support for some of this work from the NASA Origins of Solar Systems grant NNX12AJ04G.  T.B.~acknowledges funding from the European Research Council (ERC) under the European Union’s Horizon 2020 research and innovation programme under grant agreement No 714769.  The Submillimeter Array is a joint project between the Smithsonian Astrophysical Observatory and the Academia Sinica Institute of Astronomy and Astrophysics and is funded by the Smithsonian Institution and the Academia Sinica.  This work has made use of data from the European Space Agency (ESA) mission {\it Gaia} (\url{http://www.cosmos.esa.int/gaia}), processed by the {\it Gaia} Data Processing and Analysis Consortium (DPAC, \url{http://www.cosmos.esa.int/web/gaia/dpac/consortium}). Funding for the DPAC has been provided by national institutions, in particular the institutions participating in the {\it Gaia} Multilateral Agreement.

\bibliography{references}

\clearpage

\begin{deluxetable}{llcc}
\tabletypesize{\footnotesize}
\tablewidth{0pt}
\tablecolumns{5}
\tablecaption{Observations Log\label{table:obslog}}
\tablehead{
\colhead{Target} & \colhead{UT Date} & \colhead{Config.} & \colhead{Ref} \\
\colhead{(1)} & \colhead{(2)} & \colhead{(3)} & \colhead{(4)}
}
\startdata
LkH$\alpha$ 330   & 2006 Nov 19 & V  & 1   \\
                  & 2010 Nov 3  & C  & 2   \\
FN Tau            & 2008 Jul 11 & E  & 3   \\
IRAS 04125+2902   & 2011 Aug 19 & E  & 4   \\
                  & 2011 Sep 8  & V  & 4   \\
                  & 2011 Oct 26 & C  & 4   \\
CY Tau            & 2013 Sep 18 & E  & 5   \\
                  & 2013 Sep 20 & E  & 5   \\
                  & 2013 Sep 23 & E  & 5   \\ 
V410 X-ray 2      & 2011 Aug 19 & E  & 5   \\
                  & 2011 Sep 8  & V  & 5   \\
                  & 2011 Oct 26 & C  & 6   \\
IP Tau            & 2014 Aug 28 & E  & 5   \\
                  & 2014 Sep 16 & C  & 5   \\
                  & 2014 Oct 23 & C  & 5   \\
                  & 2014 Oct 30 & C  & 5   \\
                  & \dt{2015 Jan 19} & \dt{V}  & \dt{5}   \\
IQ Tau            & 2014 Aug 28 & E  & 5   \\ 
                  & 2014 Sep 12 & C  & 5   \\
                  & 2014 Oct 29 & C  & 5   \\
                  & \dt{2014 Dec 21} & \dt{V}  & \dt{5}   \\
\phm{IQ} $\vdots$ & \phm{2014} $\vdots$ & $\vdots$ & $\vdots$ \\
\enddata
\tablecomments{Table 1 is published in its entirety in the machine-readable format.
      A portion is shown here for guidance regarding its form and content.  The SMA configurations are S = sub-compact (6--70\,m baselines), C = compact (9--75\,m baselines), E = extended (17--225\,m baselines), and V = very extended (24--509\,m baselines).  References (Col.~4): 1 = \citet{brown08}, 2 = \citet{andrews11}, 3 = \citet{momose10}, 4 = \citet{espaillat15}, 5 = this paper, 6 = \citet{andrews13}, 7 = \citet{harris12}, 8 = \citet{hughes09}, 9 = \citet{isella10}, 10 = \citet{trotta13}, 11 = \citet{qi08}, 12 = \citet{hughes09b}, 13 = \citet{andrews12}, 14 = \citet{brown09}, 15 = \citet{mathews12}, 16 = \citet{andrews10b}, 17 = \citet{andrews09}, 18 = \citet{aw07a}, 19 = \citet{andrews08}, 20 = \citet{cieza12}, 21 = \citet{qi11}.}
\end{deluxetable}

\newpage
\begin{deluxetable*}{lcccccc}
\tabletypesize{\footnotesize}
\tablewidth{0pt}
\tablecolumns{7}
\tablecaption{Imaging Parameters\label{table:images}}
\tablehead{
\colhead{Target} & \colhead{R.A. (J2000)} & \colhead{Decl. (J2000)} & \colhead{mean $\nu$} &
\colhead{beam} & \colhead{beam PA} & \colhead{RMS} \\
\colhead{} & \colhead{[$^{\rm h}$ $^{\rm m}$ $^{\rm s}$]} & \colhead{[\degr\ \arcmin\ \arcsec]} & \colhead{[GHz]} & 
\colhead{[\arcsec\ $\times$ \arcsec]} & \colhead{[\degr]} & \colhead{[mJy beam$^{-1}$]}
}
\startdata
LkH$\alpha$ 330   & 03:45:48.30 & +32:24:11.99 & 340.6 & $0.37\times0.32$ & \phn82    & 2.5 \\
FN Tau            & 04:14:14.59 & +28:27:58.00 & 339.7 & $0.78\times0.70$ & \phn74    & 1.0 \\
IRAS 04125+2902   & 04:15:42.79 & +29:09:59.70 & 340.9 & $0.33\times0.26$ & \phn51    & 1.3 \\
CY Tau            & 04:17:33.73 & +28:20:46.90 & 337.1 & $0.75\times0.58$ & 104       & 1.8 \\
V410 X-ray 2      & 04:18:34.45 & +28:30:30.20 & 340.9 & $0.65\times0.56$ & \phn70    & 1.1 \\
IP Tau            & 04:24:57.08 & +27:11:56.50 & \dt{340.3} & \dt{$0.83\times0.78$} & \dt{\phn78}    & \dt{1.3} \\
IQ Tau            & 04:29:51.56 & +26:06:44.90 & \dt{340.2} & \dt{$0.50\times0.42$} & \dt{\phn64}    & \dt{1.1} \\
UX Tau A          & 04:30:03.99 & +18:13:49.40 & 340.2 & $0.39\times0.30$ & \phn55    & 1.3 \\
DK Tau A          & 04:30:44.20 & +26:01:24.80 & 341.8 & $0.43\times0.34$ & \phn69    & 1.1 \\
Haro 6-13         & 04:32:15.41 & +24:28:59.80 & 337.1 & $0.78\times0.58$ & 103       & 3.0 \\
MHO 6             & 04:32:22.11 & +18:27:42.64 & \dt{340.2} & \dt{$0.86\times0.80$} & \dt{\phn86}    & \dt{1.4} \\
FY Tau            & 04:32:30.58 & +24:19:57.28 & \dt{340.2} & \dt{$0.64\times0.56$} & \dt{\phn79}    & \dt{1.2} \\
UZ Tau E          & 04:32:43.04 & +25:52:31.10 & 335.8 & $0.43\times0.30$ & \phn49    & 1.5 \\
J04333905+2227207 & 04:33:39.05 & +22:27:20.79 & \dt{339.7} & \dt{$0.41\times0.31$} & \dt{\phn67}    & \dt{0.7} \\
J04334465+2615005 & 04:33:44.65 & +26:15:00.53 & \dt{340.3} & \dt{$0.61\times0.52$} & \dt{\phn76}    & \dt{1.2} \\
DM Tau            & 04:33:48.73 & +18:10:09.90 & 341.7 & $0.69\times0.49$ & \phn33    & 1.2 \\
CI Tau            & 04:33:52.00 & +22:50:30.20 & 337.2 & $0.79\times0.58$ & 105       & 2.3 \\
DN Tau            & 04:35:27.37 & +24:14:58.93 & \dt{340.0} & \dt{$0.60\times0.31$} & \dt{\phn93}    & \dt{1.1} \\
HP Tau            & 04:35:52.78 & +22:54:23.11 & \dt{340.2} & \dt{$0.48\times0.40$} & \dt{\phn66}    & \dt{1.2} \\
DO Tau            & 04:38:28.58 & +26:10:49.40 & 337.1 & $0.77\times0.58$ & 104       & 2.6 \\
LkCa 15           & 04:39:17.78 & +22:21:03.50 & 340.2 & $0.41\times0.32$ & \phn64    & 1.0 \\
IRAS 04385+2550   & 04:41:38.82 & +25:56:26.75 & \dt{340.2} & \dt{$0.59\times0.48$} & \dt{\phn74}    & \dt{1.1} \\
GO Tau            & 04:43:03.09 & +25:20:18.75 & \dt{340.2} & \dt{$0.56\times0.44$} & \dt{\phn77}    & \dt{1.2} \\
GM Aur            & 04:55:10.98 & +30:21:59.40 & 342.3 & $0.32\times0.27$ & \phn38    & 3.8 \\
SU Aur            & 04:55:59.38 & +30:34:01.50 & 341.8 & $1.06\times0.85$ & \phn86    & 1.6 \\
V836 Tau          & 05:03:06.60 & +25:23:19.60 & 341.4 & $0.44\times0.36$ & \phn45    & 1.1 \\
MWC 758           & 05:30:27.53 & +25:19:57.10 & 340.2 & $0.86\times0.58$ & \phn91    & 1.8 \\
CQ Tau            & 05:35:58.47 & +24:44:54.10 & 339.8 & $0.94\times0.47$ & \phn92    & 2.7 \\
TW Hya            & 11:01:51.87 & -34:42:17.11 & 336.1 & $0.81\times0.59$ & 177       & 2.8 \\
SAO 206462        & 15:15:48.40 & -37:09:16.00 & 339.7 & $0.53\times0.28$ & \phn13    & 3.8 \\
IM Lup            & 15:56:09.20 & -37:56:06.20 & 340.6 & $1.24\times1.00$ & 170       & 2.8 \\
RX J1604.3$-$2130 & 16:04:21.70 & -21:30:28.40 & 339.8 & $0.47\times0.29$ & \phn29    & 1.1 \\
RX J1615.3$-$3255 & 16:15:20.20 & -32:55:05.10 & 341.3 & $0.53\times0.24$ & \phn17    & 3.0 \\
SR 4              & 16:25:56.17 & -24:20:48.40 & 339.9 & $0.50\times0.43$ & \phn15    & 2.5 \\
Elias 20          & 16:26:18.90 & -24:28:19.80 & 340.2 & $0.44\times0.29$ & \phn26    & 1.7 \\
DoAr 25           & 16:26:23.60 & -24:43:13.20 & 346.9 & $0.49\times0.37$ & \phn14    & 3.0 \\
Elias 24          & 16:26:24.07 & -24:16:13.50 & 340.0 & $0.97\times0.79$ & \phn23    & 3.5 \\
GSS 39            & 16:26:45.03 & -24:23:07.74 & 340.0 & $0.71\times0.66$ & 138       & 3.0 \\
WL 18             & 16:26:48.98 & -24:38:25.20 & 340.0 & $0.77\times0.66$ & \phn54    & 2.5 \\
SR 24 S           & 16:26:58.51 & -24:45:37.00 & 340.0 & $0.38\times0.28$ & \phn\phn9 & 3.0 \\
SR 21             & 16:27:10.30 & -24:19:12.00 & 340.0 & $0.41\times0.29$ & \phn19    & 2.5 \\
DoAr 33           & 16:27:39.02 & -23:58:18.70 & 339.9 & $0.52\times0.43$ & \phn67    & 2.3 \\
WSB 52            & 16:27:39.43 & -24:39:15.50 & 339.9 & $0.48\times0.38$ & \phn31    & 2.3 \\
WSB 60            & 16:28:16.50 & -24:37:58.00 & 339.9 & $0.63\times0.41$ & \phn24    & 2.5 \\
SR 13             & 16:28:45.27 & -24:28:19.00 & 340.0 & $0.63\times0.52$ & \phn37    & 3.0 \\
DoAr 44           & 16:31:33.50 & -24:28:37.30 & 339.9 & $0.63\times0.41$ & \phn24    & 2.6 \\
RX J1633.9$-$2422 & 16:33:55.60 & -24:42:05.00 & 334.9 & $0.37\times0.29$ & \phn\phn8 & 1.3 \\
WaOph 6           & 16:48:45.63 & -14:16:36.00 & 345.2 & $0.54\times0.40$ & \phn34    & 2.8 \\
AS 209            & 16:49:15.30 & -14:22:08.57 & 349.1 & $0.61\times0.46$ & \phn33    & 3.3 \\
HD 163296         & 17:56:21.30 & -21:57:22.20 & 341.0 & $1.71\times1.27$ & \phn47    & 10  \\
\enddata
\end{deluxetable*}

\newpage
\begin{deluxetable*}{lccccccc|c}
\tabletypesize{\scriptsize}
\tablecolumns{10}
\tablecaption{Surface Brightness Profile Parameters\label{table:SBpars}}
\tablehead{
\colhead{Name} & \colhead{\dt{$F_\nu$} [Jy]} & \colhead{$\varrho_t$ [\arcsec]} & \colhead{$\log{\alpha}$} & \colhead{$\beta$} & \colhead{$\gamma$} & \colhead{$i$ [\degr]} & \colhead{$\varphi$ [\degr]} &  \colhead{$\varrho_{\rm eff}$ [\arcsec]}
}
\startdata 
LkH$\alpha$ 330   & 0.208 $^{+0.004}_{-0.002}$  & 0.38 $^{+0.45}_{-0.26}$  & 0.83 $^{+0.90}_{-0.10}$
                  & 5.2 $^{+2.7}_{-0.3}$        & -1.7 $^{+0.3}_{-1.0}$    & 39 $^{+3\phn}_{-5\phn}$
                  & \phn76 $^{+7\phn}_{-5\phn}$ & 0.44 $^{+0.03}_{-0.01}$  \\
FN Tau            & 0.044 $^{+0.002}_{-0.002}$  & $<0.32$                  & $p(\log{\alpha}); \downarrow$
                  & $>3.5$                      & $p(\gamma); \uparrow$    & $p(i)$
                  & $p(\varphi)$                & $<0.17$                  \\
IRAS 04125+2902   & 0.042 $^{+0.001}_{-0.001}$  & 0.27 $^{+0.03}_{-0.02}$  & $>0.74$
                  & 6.8 $^{+2.4}_{-1.0}$        & -2.5 $^{+1.1}_{-0.3}$    & 26 $^{+7\phn}_{-13}$
                  & 155 $^{+41}_{-26}$          & 0.30 $^{+0.02}_{-0.01}$  \\
CY Tau            & 0.207 $^{+0.012}_{-0.005}$  & 0.26 $^{+0.16}_{-0.01}$  & $p(\log{\alpha}); \uparrow$ 
                  & 4.3 $^{+3.7}_{-0.2}$        & \phd0.3 $^{+0.1}_{-2.6}$ & 27 $^{+3\phn}_{-4\phn}$
                  & 136 $^{+8\phn}_{-8\phn}$    & 0.36 $^{+0.02}_{-0.02}$  \\
V410 X-ray 2      & 0.026 $^{+0.002}_{-0.001}$  & 0.14 $^{+0.38}_{-0.04}$  & $p(\log{\alpha}); \downarrow$
                  & $p(\beta); \downarrow$      & $p(\gamma); \uparrow$    & 62 $^{+10}_{-27}$
                  & 106 $^{+19}_{-20}$          & 0.16 $^{+0.05}_{-0.02}$  \\
IP Tau            & \dt{0.030 $^{+0.002}_{-0.001}$}  & \dt{0.25 $^{+0.09}_{-0.04}$}  & \dt{$p(\log{\alpha}); \uparrow$}
                  & \dt{$>3.0$}                      & \dt{-1.2 $^{+1.4}_{-1.3}$}    & \dt{41 $^{+12}_{-20}$}
                  & \dt{\phn40 $^{+93}_{-17}$}       & \dt{0.27 $^{+0.05}_{-0.04}$}  \\
IQ Tau            & \dt{0.163 $^{+0.004}_{-0.002}$}  & \dt{0.52 $^{+0.39}_{-0.04}$}  & \dt{0.40 $^{+0.57}_{-0.13}$} 
                  & \dt{4.4 $^{+4.5}_{-0.1}$}        & \dt{\phd0.6 $^{+0.1}_{-0.5}$} & \dt{60 $^{+1\phn}_{-1\phn}$}
                  & \dt{\phn43 $^{+2\phn}_{-2\phn}$} & \dt{0.55 $^{+0.01}_{-0.02}$}  \\
UX Tau A          & 0.147 $^{+0.001}_{-0.002}$  & 0.27 $^{+0.01}_{-0.01}$  & $>1.25$
                  & $>8.0$                      & -2.9 $^{+0.3}_{-0.4}$    & 39 $^{+2\phn}_{-2\phn}$
                  & 167 $^{+3\phn}_{-2\phn}$    & 0.28 $^{+0.01}_{-0.01}$  \\
DK Tau A          & 0.071 $^{+0.004}_{-0.006}$  & 0.08 $^{+0.11}_{-0.01}$  & $p(\log{\alpha}); \downarrow$ 
                  & 2.6 $^{+2.0}_{-0.1}$        & \phd0.9 $^{+0.1}_{-2.8}$ & 26 $^{+7\phn}_{-12}$
                  & 106 $^{+24}_{-32}$          & 0.32 $^{+0.15}_{-0.02}$  \\
Haro 6-13         & 0.441 $^{+0.069}_{-0.022}$  & 0.06 $^{+0.08}_{-0.01}$  & $p(\log{\alpha}); \uparrow$
                  & 2.3 $^{+0.3}_{-0.1}$        & $p(\gamma); \uparrow$    & 42 $^{+3\phn}_{-3\phn}$
                  & 145 $^{+6\phn}_{-6\phn}$    & 0.48 $^{+0.43}_{-0.05}$  \\
MHO 6             & \dt{0.047 $^{+0.002}_{-0.002}$}  & \dt{0.23 $^{+0.17}_{-0.05}$}   & \dt{$p(\log{\alpha}); \downarrow$}
                  & \dt{$p(\beta); \uparrow$}        & \dt{$p(\gamma); \uparrow$}    & \dt{78 $^{+7\phn}_{-14}$}
                  & \dt{110 $^{+11}_{-8\phn}$}          & \dt{0.26 $^{+0.06}_{-0.04}$}  \\
FY Tau            & \dt{0.023 $^{+0.002}_{-0.001}$}  & \dt{$<$0.50}                  & \dt{$p(\log{\alpha}); \downarrow$}
                  & \dt{$p(\beta)$; $\uparrow$}                      & \dt{$p(\gamma); \uparrow$}    & \dt{$>25$}
                  & \dt{134 $^{+20}_{-62}$}          & \dt{0.07 $^{+0.04}_{-0.04}$}                  \\
UZ Tau E          & 0.401 $^{+0.008}_{-0.007}$  & 0.55 $^{+0.01}_{-0.02}$  & 0.83 $^{+0.13}_{-0.08}$
                  & 3.3 $^{+0.2}_{-0.1}$        & \phd0.6 $^{+0.1}_{-0.1}$ & 55 $^{+1\phn}_{-1\phn}$
                  & \phn90 $^{+1\phn}_{-1\phn}$ & 0.80 $^{+0.03}_{-0.02}$  \\
J04333905+2227207 & \dt{0.076 $^{+0.002}_{-0.001}$}  & \dt{0.58 $^{+0.14}_{-0.04}$}  & \dt{0.66 $^{+0.83}_{-0.09}$}
                  & \dt{$p(\beta); \uparrow$}        & \dt{\phd0.3 $^{+0.1}_{-0.3}$}    & \dt{79 $^{+1\phn}_{-1\phn}$}
                  & \dt{115 $^{+1\phn}_{-2\phn}$}    & \dt{0.56 $^{+0.02}_{-0.02}$}  \\
J04334465+2615005 & \dt{0.043 $^{+0.002}_{-0.002}$}  & \dt{0.46 $^{+0.20}_{-0.15}$}  & \dt{$p(\log{\alpha})$; $\downarrow$}
                  & \dt{$>4.1$}                      & \dt{\phd1.2 $^{+0.1}_{-1.2}$}              & \dt{72 $^{+11}_{-17}$}
                  & \dt{158 $^{+5\phn}_{-7\phn}$}       & \dt{0.34 $^{+0.04}_{-0.05}$}  \\
DM Tau            & 0.254 $^{+0.014}_{-0.008}$  & 0.15 $^{+0.07}_{-0.04}$  & $<0.31$           
                  & 3.0 $^{+0.3}_{-0.3}$        & -1.4 $^{+0.6}_{-1.2}$    & 34 $^{+2\phn}_{-2\phn}$
                  & 158 $^{+5\phn}_{-5\phn}$    & 0.92 $^{+0.15}_{-0.06}$  \\
CI Tau            & 0.440 $^{+0.050}_{-0.014}$  & 1.45 $^{+0.69}_{-0.30}$  & 0.38 $^{+0.12}_{-0.20}$ 
                  & $>4.1$                      & \phd0.8 $^{+0.1}_{-0.2}$ & 44 $^{+1\phn}_{-3\phn}$ 
                  & \phn11 $^{+3\phn}_{-2\phn}$ & 0.89 $^{+0.13}_{-0.04}$  \\
DN Tau            & \dt{0.179 $^{+0.005}_{-0.003}$}  & \dt{0.35 $^{+0.29}_{-0.02}$}  & \dt{$p(\log{\alpha}); \downarrow$} 
                  & \dt{$p(\beta); \downarrow$}      & \dt{\phd0.8 $^{+0.1}_{-0.8}$} & \dt{28 $^{+2\phn}_{-8\phn}$}
                  & \dt{\phn80 $^{+11}_{-9\phn}$}       & \dt{0.38 $^{+0.02}_{-0.01}$}  \\
HP Tau            & \dt{0.106 $^{+0.001}_{-0.002}$}  & \dt{0.44 $^{+0.05}_{-0.04}$}                  & \dt{$p(\log{\alpha}); \uparrow$}
                  & $>3.8$                      & \dt{\phd1.6 $^{+0.1}_{-0.1}$}    & \dt{64 $^{+2\phn}_{-3\phn}$}
                  & \dt{\phn70 $^{+3\phn}_{-3\phn}$}       & \dt{0.22 $^{+0.02}_{-0.01}$}                  \\
DO Tau            & 0.250 $^{+0.012}_{-0.003}$  & 0.12 $^{+0.28}_{-0.01}$  & $p(\log{\alpha}); \downarrow$
                  & $3.7^{+4.8}_{-0.7}$         & $p(\gamma); \uparrow$    & 37 $^{+4\phn}_{-7\phn}$
                  & 159 $^{+7\phn}_{-10}$       & 0.19 $^{+0.01}_{-0.01}$  \\
LkCa 15           & 0.396 $^{+0.005}_{-0.004}$  & 0.62 $^{+0.05}_{-0.02}$  & 0.62 $^{+0.08}_{-0.10}$
                  & 5.3 $^{+1.0}_{-0.3}$        & -1.6 $^{+0.2}_{-0.4}$    & 51 $^{+1\phn}_{-1\phn}$
                  & \phn62 $^{+1\phn}_{-1\phn}$ & 0.77 $^{+0.01}_{-0.01}$  \\
IRAS 04385+2550   & \dt{0.061 $^{+0.004}_{-0.003}$}  & \dt{$p(\varrho_t)$; $\downarrow$}  & \dt{$p(\log{\alpha})$; $\downarrow$}
                  & \dt{$p(\beta)$}                      & \dt{$>$1.4}    & \dt{$p(i)$}
                  & \dt{148 $^{+6\phn}_{-19}$}       & \dt{0.10 $^{+0.08}_{-0.01}$}  \\
GO Tau            & \dt{0.179 $^{+0.019}_{-0.007}$}  & \dt{0.89 $^{+2.73}_{-0.05}$}  & \dt{0.20 $^{+1.05}_{-0.08}$}
                  & \dt{$p(\beta)$; $\downarrow$}        & \dt{1.1 $^{+0.1}_{-0.1}$}     & \dt{53 $^{+2\phn}_{-3\phn}$}
                  & \dt{\phn26 $^{+4\phn}_{-3\phn}$} & \dt{1.16 $^{+0.23}_{-0.07}$}  \\
GM Aur            & 0.632 $^{+0.011}_{-0.008}$  & 0.92 $^{+0.52}_{-0.27}$  & $<0.26$
                  & 6.7 $^{+2.3}_{-1.1}$        & -1.1 $^{+0.4}_{-0.7}$    & 55 $^{+1\phn}_{-1\phn}$
                  & \phn64 $^{+1\phn}_{-1\phn}$ & 0.87 $^{+0.03}_{-0.02}$  \\
SU Aur            & 0.052 $^{+0.002}_{-0.001}$  & $<0.27$                  & $p(\log{\alpha}); \downarrow$
                  & $>3.5$                      & $p(\gamma); \uparrow$    & $p(i)$
                  & \phn79 $^{+57}_{-42}$       & $<0.16$                  \\
V836 Tau          & 0.061 $^{+0.002}_{-0.002}$  & 0.13 $^{+0.05}_{-0.01}$  & $p(\log{\alpha}); \uparrow$
                  & $>4.7$                      & $p(\gamma)$              & 61 $^{+11}_{-8\phn}$
                  & 137 $^{+9\phn}_{-6\phn}$    & 0.13 $^{+0.02}_{-0.01}$  \\
MWC 758           & 0.177 $^{+0.002}_{-0.001}$  & 0.50 $^{+0.01}_{-0.01}$  & $>1.3$
                  & $>7.9$                      & -3.0 $^{+0.4}_{-0.4}$    & 40 $^{+1\phn}_{-1\phn}$
                  & 168 $^{+1\phn}_{-1\phn}$    & 0.53 $^{+0.01}_{-0.01}$  \\
CQ Tau            & 0.444 $^{+0.013}_{-0.009}$  & 0.34 $^{+0.03}_{-0.01}$  & $>0.87$
                  & $>6.6$                      & -2.3 $^{+1.1}_{-0.5}$    & 36 $^{+3\phn}_{-3\phn}$
                  & \phn53 $^{+6\phn}_{-6\phn}$ & 0.36 $^{+0.01}_{-0.02}$  \\
TW Hya            & 1.311 $^{+0.003}_{-0.002}$  & 0.98 $^{+0.01}_{-0.02}$  & 0.95 $^{+0.05}_{-0.03}$
                  & $>8.6$                      & \phd0.6 $^{+0.1}_{-0.1}$ & \phn7 $^{+1\phn}_{-1\phn}$
                  & 155 $^{+1\phn}_{-1\phn}$    & 0.77 $^{+0.01}_{-0.01}$  \\
SAO 206462        & 0.596 $^{+0.024}_{-0.022}$  & 0.57 $^{+0.02}_{-0.04}$  & 0.88 $^{+0.59}_{-0.08}$
                  & $>5.9$                      & -1.0 $^{+0.1}_{-0.7}$    & 27 $^{+3\phn}_{-4\phn}$
                  & \phn33 $^{+7\phn}_{-9\phn}$ & 0.56 $^{+0.02}_{-0.01}$  \\
IM Lup            & 0.587 $^{+0.007}_{-0.006}$  & 2.25 $^{+0.48}_{-0.42}$  & 0.32 $^{+0.12}_{-0.07}$
                  & $>5.5$                      & \phd0.9 $^{+0.1}_{-0.1}$ & 49 $^{+1\phn}_{-2\phn}$
                  & 142 $^{+2\phn}_{-3\phn}$    & 1.11 $^{+0.04}_{-0.02}$  \\
RX J1604.3$-$2130 & 0.194 $^{+0.009}_{-0.006}$  & 0.63 $^{+0.01}_{-0.01}$  & $>1.2$
                  & 5.8 $^{+0.5}_{-0.4}$        & -2.5 $^{+0.3}_{-0.3}$    & \phn6 $^{+1\phn}_{-1\phn}$
                  & \phn77 $^{+1\phn}_{-1\phn}$ & 0.72 $^{+0.01}_{-0.02}$  \\
RX J1615.3$-$3255 & 0.434 $^{+0.005}_{-0.003}$  & 1.39 $^{+0.24}_{-0.51}$  & $<0.31$
                  & $>5.1$                      & -0.1 $^{+0.3}_{-0.3}$    & 44 $^{+2\phn}_{-2\phn}$
                  & 149 $^{+3\phn}_{-3\phn}$    & 0.73 $^{+0.02}_{-0.02}$  \\
SR 4              & 0.148 $^{+0.003}_{-0.003}$  & 0.19 $^{+0.05}_{-0.01}$  & $>0.45$ 
                  & $>5.2$                      & -0.9 $^{+0.9}_{-1.5}$    & 43 $^{+3\phn}_{-6\phn}$
                  & \phn27 $^{+8\phn}_{-6\phn}$ & 0.21 $^{+0.01}_{-0.01}$  \\ 
Elias 20          & 0.264 $^{+0.006}_{-0.004}$  & 0.41 $^{+0.05}_{-0.05}$  & $>0.43$
                  & 7.0 $^{+2.1}_{-1.5}$        & \phd1.0 $^{+0.1}_{-0.2}$ & 54 $^{+2\phn}_{-2\phn}$
                  & 162 $^{+2\phn}_{-2\phn}$    & 0.32 $^{+0.02}_{-0.01}$  \\
DoAr 25           & 0.565 $^{+0.008}_{-0.008}$  & 1.18 $^{+0.09}_{-0.18}$  & 0.58 $^{+0.14}_{-0.07}$ 
                  & $>5.5$                      & \phd0.5 $^{+0.1}_{-0.1}$ & 63 $^{+1\phn}_{-1\phn}$   
                  & 109 $^{+1\phn}_{-1\phn}$    & 0.85 $^{+0.02}_{-0.02}$  \\
Elias 24          & 0.912 $^{+0.016}_{-0.011}$  & 0.82 $^{+0.09}_{-0.03}$  & $>0.76$
                  & 5.9 $^{+2.2}_{-0.5}$        & \phd1.1 $^{+0.1}_{-0.1}$ & 21 $^{+3\phn}_{-5\phn}$
                  & \phn62 $^{+13}_{-11}$       & 0.68 $^{+0.01}_{-0.02}$  \\
GSS 39            & 0.654 $^{+0.012}_{-0.010}$  & 1.61 $^{+0.05}_{-0.05}$  & $>1.04$
                  & $>7.2$                      & \phd1.1 $^{+0.1}_{-0.1}$ & 55 $^{+1\phn}_{-1\phn}$
                  & 120 $^{+1\phn}_{-1\phn}$    & 1.19 $^{+0.02}_{-0.09}$  \\
WL 18             & 0.051 $^{+0.004}_{-0.002}$  & 0.13 $^{+0.08}_{-0.03}$  & $p(\log{\alpha}); \uparrow$
                  & $>3.8$                      & $p(\gamma)$              & 60 $^{+17}_{-29}$
                  & \phn68 $^{+54}_{-23}$       & 0.13 $^{+0.04}_{-0.02}$  \\
SR 24 S           & 0.509 $^{+0.006}_{-0.006}$  & 0.43 $^{+0.05}_{-0.02}$  & 0.65 $^{+0.87}_{-0.08}$
                  & 6.2 $^{+2.9}_{-0.4}$        & -0.4 $^{+0.1}_{-1.1}$    & 46 $^{+2\phn}_{-2\phn}$ 
                  & \phn23 $^{+2\phn}_{-2\phn}$ & 0.45 $^{+0.08}_{-0.08}$  \\
SR 21             & 0.405 $^{+0.003}_{-0.003}$  & 0.44 $^{+0.02}_{-0.02}$  & $>0.95$
                  & $>6.9$                      & -0.9 $^{+0.2}_{-0.6}$    & 18 $^{+5\phn}_{-9\phn}$
                  & \phn10 $^{+45}_{-49}$       & 0.43 $^{+0.02}_{-0.01}$  \\
DoAr 33           & 0.070 $^{+0.008}_{-0.004}$  & 0.19 $^{+0.29}_{-0.04}$  & $p(\log{\alpha}); \downarrow$
                  & $p(\beta); \uparrow$        & $p(\gamma); \uparrow$    & 68 $^{+11}_{-27}$
                  & \phn76 $^{+21}_{-19}$       & 0.21 $^{+0.13}_{-0.03}$  \\
WSB 52            & 0.154 $^{+0.016}_{-0.005}$  & 0.93 $^{+1.26}_{-0.24}$  & $p(\log{\alpha}); \downarrow$
                  & $>3.1$                      & \phd1.6 $^{+0.1}_{-0.3}$ & 47 $^{+9\phn}_{-24}$
                  & 165 $^{+35}_{-26}$          & 0.42 $^{+0.12}_{-0.06}$  \\
WSB 60            & 0.258 $^{+0.015}_{-0.006}$  & 0.26 $^{+0.01}_{-0.01}$  & 0.37 $^{+1.25}_{-0.02}$
                  & 3.5 $^{+3.2}_{-0.2}$        & -0.3 $^{+0.1}_{-2.1}$    & 32 $^{+4\phn}_{-12}$
                  & 121 $^{+17}_{-17}$          & 0.34 $^{+0.03}_{-0.02}$  \\
SR 13             & 0.153 $^{+0.004}_{-0.003}$  & 0.31 $^{+0.03}_{-0.03}$  & $>0.67$
                  & $>5.1$                      & -2.4 $^{1.3}_{-0.4}$     & 54 $^{+4\phn}_{-6\phn}$
                  & \phn72 $^{+6\phn}_{-7\phn}$ & 0.32 $^{+0.03}_{-0.01}$  \\
DoAr 44           & 0.211 $^{+0.006}_{-0.005}$  & 0.38 $^{+0.05}_{-0.03}$  & 0.69 $^{+0.87}_{-0.06}$
                  & 5.6 $^{+3.3}_{-0.4}$        & -1.8 $^{+0.7}_{-0.8}$    & 16 $^{+10}_{-6\phn}$
                  & \phn55 $^{+92}_{-27}$       & 0.42 $^{+0.02}_{-0.01}$  \\
RX J1633.9$-$2422 & 0.225 $^{+0.011}_{-0.005}$  & 0.34 $^{+0.01}_{-0.03}$  & 0.97 $^{+0.80}_{-0.07}$
                  & 6.7 $^{+2.3}_{-0.7}$        & -1.5 $^{+0.2}_{-0.9}$    & 49 $^{+1\phn}_{-1\phn}$
                  & 83 $^{+2\phn}_{-1\phn}$     & 0.35 $^{+0.01}_{-0.01}$  \\
WaOph 6           & 0.445 $^{+0.024}_{-0.022}$  & 0.47 $^{+0.09}_{-0.03}$  & $>0.33$
                  & 2.9 $^{+0.5}_{-0.2}$        & \phd1.1 $^{+0.1}_{-0.1}$ & 41 $^{+4\phn}_{-3\phn}$
                  & 173 $^{+10}_{-3\phn}$       & 0.68 $^{+0.15}_{-0.05}$  \\  
AS 209            & 0.604 $^{+0.007}_{-0.010}$  & 1.65 $^{+0.13}_{-0.50}$  & $<0.16$
                  & $>6.4$                      & -0.2 $^{+0.3}_{-0.1}$    & 31 $^{+3\phn}_{-5\phn}$
                  & \phn73 $^{+6\phn}_{-9\phn}$ & 0.70 $^{+0.02}_{-0.02}$  \\
HD 163296         & 1.822 $^{+0.004}_{-0.006}$  & 1.60 $^{+0.04}_{-0.07}$  & 0.54 $^{+0.04}_{-0.04}$
                  & $>8.6$                      & \phd0.8 $^{+0.1}_{-0.1}$ & 47 $^{+1\phn}_{-1\phn}$
                  & 133 $^{+1\phn}_{-1\phn}$    & 0.96 $^{+0.01}_{-0.01}$  \\
\enddata
\tablecomments{Quoted values are the peaks of the posterior distributions, with uncertainties that represent the 68\%\ confidence interval.  Limits represent the 95\%\ confidence boundary.  A note of ``$p(X)$" means the posterior is identical to the prior at $<$68\%\ confidence; ``$p(X); \downarrow$"  or ``$p(X); \uparrow$" means the posterior is consistent with the prior at 95\%\ confidence, but has a marginal preference toward the lower or upper bound of the prior on $X$, respectively.  Note that \dt{the $F_\nu$ posterior does not include systematic flux calibration uncertainties (see Sect.~\ref{sec:convert}) and that} the $\varrho_{\rm eff}$ posterior is not {\it directly} inferred, but rather constructed from the joint posterior on \{$F_{\nu}$, $\varrho_t$, $\log{\alpha}$, $\beta$, $\gamma$\}.}
\end{deluxetable*}

\newpage
\begin{deluxetable*}{lcc|cccc}
\tabletypesize{\scriptsize}
\tablecolumns{7}
\tablecaption{Size--Luminosity Relation and Associated Parameters\label{table:sizelum}}
\tablehead{
\colhead{Name} & 
\colhead{$\log{R_{\rm eff}/{\rm AU}}$} & 
\colhead{$\log{L_{\rm mm}/{\rm Jy}}$} & 
\colhead{$\varpi$ [mas]} & 
\colhead{$d$ [pc]} & 
\colhead{$L_\ast$ [$L_\odot$]} & 
\colhead{$\langle T_d \rangle$ [K]}
}
\startdata 
LkH$\alpha$ 330   & 2.13 $^{+0.09}_{-0.06}$ & -0.01 $^{+0.17}_{-0.10}$ 
                  & \phn$3.38\pm0.50$ & 297 $^{+72\phn}_{-29\phn}$ 
                  & 19.6 $^{+14.5}_{-3.7\phn}$ & 30 $^{+7\phn}_{-5\phn}$\\
FN Tau            & $<1.43$                 & -1.40 $^{+0.12}_{-0.09}$
                  & \phn$7.62\pm0.86$ & 131 $^{+22\phn}_{-10\phn}$ 
                  & 0.65 $^{+0.32}_{-0.13}$ & 30 $^{+35}_{-4\phn}$ \\       
IRAS 04125+2902   & 1.60 $^{+0.07}_{-0.05}$ & -1.42 $^{+0.12}_{-0.08}$ 
                  & \phn$7.62\pm0.86$ & 131 $^{+22\phn}_{-10\phn}$ 
                  & 0.25 $^{+0.12}_{-0.05}$ & 12 $^{+3\phn}_{-2\phn}$ \\      
CY Tau            & 1.69 $^{+0.06}_{-0.05}$ & -0.72 $^{+0.12}_{-0.09}$ 
                  & \phn$7.62\pm0.86$ & 131 $^{+22\phn}_{-11\phn}$ 
                  & 0.32 $^{+0.15}_{-0.07}$ & 12 $^{+3\phn}_{-2\phn}$ \\
V410 X-ray 2      & 1.35 $^{+0.11}_{-0.10}$ & -1.62 $^{+0.12}_{-0.09}$
                  & \phn$7.62\pm0.86$ & 132 $^{+21\phn}_{-11\phn}$ 
                  & 2.09 $^{+1.07}_{-0.40}$ & 34 $^{+14}_{-6\phn}$ \\
IP Tau            & \dt{1.56 $^{+0.10}_{-0.08}$} & \dt{-1.56 $^{+0.15}_{-0.10}$}
                  & \phn$7.63\pm1.01$ & 131 $^{+27\phn}_{-12\phn}$ 
                  & 0.37 $^{+0.21}_{-0.08}$ & \dt{15 $^{+3\phn}_{-3\phn}$} \\
IQ Tau            & \dt{1.90 $^{+0.08}_{-0.05}$} & \dt{-0.76 $^{+0.16}_{-0.10}$}
                  & \phn$7.07\pm0.98$ & 143 $^{+30\phn}_{-15\phn}$  
                  & 0.73 $^{+0.48}_{-0.14}$ & \dt{14 $^{+3\phn}_{-2\phn}$} \\       
UX Tau A          & 1.65 $^{+0.03}_{-0.03}$ & -0.73 $^{+0.06}_{-0.05}$
                  & \phn$6.33\pm0.40$ & 158 $^{+12\phn}_{-8\phn\phn}$ 
                  & 2.18 $^{+0.66}_{-0.42}$ & 21 $^{+3\phn}_{-4\phn}$ \\   
DK Tau A          & 1.71 $^{+0.15}_{-0.09}$ & -1.13 $^{+0.16}_{-0.10}$
                  & \phn$7.07\pm0.98$ & 142 $^{+31\phn}_{-14\phn}$ 
                  & 1.19 $^{+0.77}_{-0.24}$ & 22 $^{+6\phn}_{-4\phn}$ \\      
Haro 6-13         & 1.97 $^{+0.21}_{-0.15}$ & -0.26 $^{+0.22}_{-0.13}$
                  & \phn$6.79\pm1.19$ & 147 $^{+50\phn}_{-15\phn}$ 
                  & 0.66 $^{+0.65}_{-0.12}$ & 17 $^{+5\phn}_{-4\phn}$ \\      
MHO 6             & \dt{1.64 $^{+0.23}_{-0.10}$} & \dt{-1.20 $^{+0.42}_{-0.15}$}
                  & \phn$6.55\pm1.58$ & 153 $^{+110}_{-16\phn}$ 
                  & 0.08 $^{+0.16}_{-0.02}$ & \dt{11 $^{+5\phn}_{-2\phn}$} \\         
FY Tau            & \dt{1.11 $^{+0.14}_{-0.41}$}            & \dt{-1.56 $^{+0.23}_{-0.12}$}
                  & \phn$6.79\pm1.19$ & 148 $^{+49\phn}_{-16\phn}$ 
                  & 1.55 $^{+1.50}_{-0.30}$ & \dt{44 $^{+36}_{-7\phn}$} \\      
UZ Tau E          & 2.06 $^{+0.09}_{-0.05}$ & -0.36 $^{+0.16}_{-0.11}$
                  & \phn$7.07\pm1.02$ & 141 $^{+34\phn}_{-13\phn}$ 
                  & 0.83 $^{+0.58}_{-0.16}$ & 12 $^{+3\phn}_{-2\phn}$\\      
J04333905+2227207 & \dt{1.95 $^{+0.15}_{-0.07}$} & \dt{-0.99 $^{+0.29}_{-0.15}$}
                  & \phn$6.55\pm1.36$ & 153 $^{+73\phn}_{-17\phn}$ 
                  & 0.07 $^{+0.10}_{-0.01}$ & \dt{\phn8 $^{+1\phn}_{-1\phn}$} \\      
J04334465+2615005 & \dt{1.69 $^{+0.10}_{-0.09}$} & \dt{-1.33 $^{+0.12}_{-0.16}$}
                  & \phn$7.07\pm1.02$ & 141 $^{+34\phn}_{-12\phn}$ 
                  & 0.24 $^{+0.17}_{-0.04}$ & \dt{14 $^{+4\phn}_{-3\phn}$} \\      
DM Tau            & 2.21 $^{+0.22}_{-0.10}$ & -0.44 $^{+0.43}_{-0.17}$
                  & \phn$6.54\pm1.62$ & 153 $^{+121}_{-16\phn}$ 
                  & 0.26 $^{+0.55}_{-0.07}$ & \phn9 $^{+4\phn}_{-1\phn}$ \\         
CI Tau            & 2.18 $^{+0.26}_{-0.08}$ & -0.22 $^{+0.35}_{-0.14}$
                  & \phn$6.54\pm1.45$ & 155 $^{+81\phn}_{-18\phn}$ 
                  & 0.93 $^{+1.66}_{-0.15}$ & 12 $^{+4\phn}_{-2\phn}$ \\      
DN Tau            & \dt{1.81 $^{+0.12}_{-0.07}$} & \dt{-0.59 $^{+0.23}_{-0.24}$}
                  & \phn$6.25\pm1.16$ & 160 $^{+60\phn}_{-17\phn}$ 
                  & 0.87 $^{+0.99}_{-0.15}$ & \dt{18 $^{+5\phn}_{-3\phn}$} \\      
HP Tau            & \dt{1.56 $^{+0.02}_{-0.03}$} & \dt{-0.85 $^{+0.01}_{-0.02}$}
                  & \phn$6.20\pm0.10$ & 161 $^{+3\phn\phn}_{-3\phn\phn}$ 
                  & 2.43 $^{+0.49}_{-0.49}$ & \dt{40 $^{+7\phn}_{-7\phn}$} \\
DO Tau            & 1.46 $^{+0.11}_{-0.07}$ & -0.52 $^{+0.21}_{-0.12}$
                  & \phn$6.79\pm1.16$ & 147 $^{+47\phn}_{-15\phn}$ 
                  & 1.33 $^{+1.25}_{-0.25}$ & 28 $^{+10}_{-5\phn}$ \\      
LkCa 15           & 2.12 $^{+0.23}_{-0.08}$ & -0.23 $^{+0.46}_{-0.15}$
                  & \phn$6.28\pm1.58$ & 161 $^{+130}_{-18\phn}$ 
                  & 0.95 $^{+2.33}_{-0.20}$ & 12 $^{+4\phn}_{-2\phn}$ \\         
IRAS 04385+2550   & \dt{1.31 $^{+0.22}_{-0.12}$} & \dt{-1.06 $^{+0.25}_{-0.13}$}
                  & \phn$6.25\pm1.16$ & 161 $^{+59\phn}_{-18\phn}$ 
                  & 0.42 $^{+0.47}_{-0.08}$ & \dt{42 $^{+15}_{-9\phn}$} \\      
GO Tau            & \dt{2.30 $^{+0.15}_{-0.07}$} & \dt{-0.59 $^{+0.25}_{-0.13}$}
                  & \phn$6.25\pm1.16$ & 160 $^{+60\phn}_{-17\phn}$ 
                  & 0.32 $^{+0.36}_{-0.06}$ & \dt{\phn8 $^{+3\phn}_{-1\phn}$} \\
GM Aur            & 2.07 $^{+0.08}_{-0.05}$ & -0.25 $^{+0.15}_{-0.09}$
                  & \phn$7.64\pm1.02$ & 131 $^{+27\phn}_{-12\phn}$ 
                  & 0.96 $^{+0.58}_{-0.20}$ & 12 $^{+2\phn}_{-3\phn}$ \\      
SU Aur            & $<1.43$                 & -1.26 $^{+0.10}_{-0.07}$
                  & \phn$7.02\pm0.67$ & 142 $^{+19\phn}_{-10\phn}$ 
                  & 9.92 $^{+4.18}_{-1.93}$ & 62 $^{+67}_{-9\phn}$ \\      
V836 Tau          & 1.24 $^{+0.06}_{-0.05}$ & -1.30 $^{+0.10}_{-0.07}$
                  & \phn$7.93\pm0.72$ & 126 $^{+15\phn}_{-9\phn\phn}$ 
                  & 0.41 $^{+0.17}_{-0.08}$ & 21 $^{+6\phn}_{-3\phn}$ \\   
MWC 758           & 1.90 $^{+0.04}_{-0.03}$ & -0.69 $^{+0.08}_{-0.05}$
                  & \phn$6.63\pm0.47$  & 150 $^{+14\phn}_{-8\phn}$ 
                  & 8.52 $^{+2.49}_{-1.84}$ & 21 $^{+4\phn}_{-4\phn}$ \\ 
CQ Tau            & 1.76 $^{+0.04}_{-0.03}$ & -0.24 $^{+0.07}_{-0.05}$
                  & \phn$6.25\pm0.40$  & 159 $^{+13\phn}_{-8\phn\phn}$ 
                  & 16.7 $^{+4.8\phn}_{-3.4\phn}$ & 31 $^{+6\phn}_{-4\phn}$ \\ 
TW Hya            & 1.66 $^{+0.01}_{-0.01}$ & -0.63 $^{+0.02}_{-0.02}$
                  & $16.80\pm0.40$     & \phn59 $^{+2\phn\phn}_{-2\phn\phn}$ 
                  & 0.33 $^{+0.08}_{-0.06}$ & \phn 9 $^{+1\phn}_{-1\phn}$ \\
SAO 206462        & 1.95 $^{+0.04}_{-0.04}$ & -0.12 $^{+0.08}_{-0.07}$
                  & \phn$6.41\pm0.54$  & 156 $^{+17\phn}_{-10\phn}$ 
                  & 9.62 $^{+3.24}_{-2.09}$ & 22 $^{+4\phn}_{-4\phn}$ \\
IM Lup            & 2.26 $^{+0.02}_{-0.03}$ & -0.11 $^{+0.08}_{-0.06}$
                  & \phn$6.20\pm0.47$  & 161 $^{+16\phn}_{-10\phn}$ 
                  & 0.90 $^{+0.28}_{-0.19}$ & \phn9 $^{+4\phn}_{-1\phn}$ \\
RX J1604.3$-$2130 & 2.03 $^{+0.08}_{-0.05}$ & -0.67 $^{+0.17}_{-0.10}$
                  & \phn$6.90\pm0.97$  & 144 $^{+34\phn}_{-13\phn}$ 
                  & 0.55 $^{+0.36}_{-0.11}$ & \phn9 $^{+2\phn}_{-1\phn}$ \\
RX J1615.3$-$3255 & 2.03 $^{+0.11}_{-0.06}$ & -0.33 $^{+0.23}_{-0.12}$
                  & \phn$7.14\pm1.26$  & 140 $^{+48\phn}_{-14\phn}$ 
                  & 0.68 $^{+0.70}_{-0.12}$ & 12 $^{+3\phn}_{-2\phn}$ \\
SR 4              & 1.45 $^{+0.03}_{-0.03}$ & -0.85 $^{+0.01}_{-0.01}$
                  & \phn$7.28\pm0.06$ & 137 $^{+1\phn\phn}_{-1\phn\phn}$ 
                  & 1.22 $^{+0.33}_{-0.25}$ & 22 $^{+4\phn}_{-4\phn}$ \\    
Elias 20          & 1.65 $^{+0.02}_{-0.02}$ & -0.59 $^{+0.01}_{-0.01}$
                  & \phn$7.28\pm0.06$ & 137 $^{+1\phn\phn}_{-1\phn\phn}$ 
                  & 2.31 $^{+0.44}_{-0.48}$ & 25 $^{+4\phn}_{-5\phn}$ \\    
DoAr 25           & 2.07 $^{+0.01}_{-0.01}$ & -0.26 $^{+0.01}_{-0.01}$
                  & \phn$7.28\pm0.06$ & 137 $^{+1\phn\phn}_{-1\phn\phn}$ 
                  & 0.96 $^{+0.20}_{-0.18}$ & 11 $^{+3\phn}_{-2\phn}$ \\
Elias 24          & 1.97 $^{+0.01}_{-0.01}$ & -0.06 $^{+0.01}_{-0.01}$
                  & \phn$7.28\pm0.06$ & 137 $^{+1\phn\phn}_{-1\phn\phn}$ 
                  & 6.16 $^{+1.24}_{-1.23}$ & 22 $^{+5\phn}_{-3\phn}$ \\   
GSS 39            & 2.21 $^{+0.01}_{-0.01}$ & -0.20 $^{+0.01}_{-0.01}$
                  & \phn$7.28\pm0.06$ & 137 $^{+1\phn\phn}_{-1\phn\phn}$ 
                  & 1.21 $^{+0.34}_{-0.24}$ & 11 $^{+2\phn}_{-2\phn}$ \\    
WL 18             & 1.29 $^{+0.07}_{-0.12}$ & -1.30 $^{+0.03}_{-0.03}$
                  & \phn$7.28\pm0.06$ & 137 $^{+1\phn\phn}_{-1\phn\phn}$ 
                  & 0.37 $^{+0.07}_{-0.08}$ & 20 $^{+8\phn}_{-3\phn}$ \\    
SR 24 S           & 1.79 $^{+0.01}_{-0.01}$ & -0.31 $^{+0.01}_{-0.01}$
                  & \phn$7.28\pm0.06$ & 137 $^{+1\phn\phn}_{-1\phn\phn}$ 
                  & 5.26 $^{+1.12}_{-1.01}$ & 22 $^{+4\phn}_{-4\phn}$ \\    
SR 21             & 1.78 $^{+0.02}_{-0.01}$ & -0.41 $^{+0.01}_{-0.01}$
                  & \phn$7.28\pm0.06$ & 137 $^{+1\phn\phn}_{-1\phn\phn}$ 
                  & 13.3 $^{+2.6\phn}_{-2.7\phn}$ & 27 $^{+5\phn}_{-5\phn}$ \\    
DoAr 33           & 1.48 $^{+0.19}_{-0.07}$ & -1.17 $^{+0.05}_{-0.03}$
                  & \phn$7.28\pm0.12$ & 137 $^{+1\phn\phn}_{-1\phn\phn}$ 
                  & 1.45 $^{+0.30}_{-0.29}$ & 23 $^{+7\phn}_{-4\phn}$ \\    
WSB 52            & 1.77 $^{+0.10}_{-0.08}$ & -0.82 $^{+0.04}_{-0.02}$
                  & \phn$7.28\pm0.06$ & 137 $^{+1\phn\phn}_{-1\phn\phn}$ 
                  & 0.72 $^{+0.15}_{-0.14}$ & 20 $^{+4\phn}_{-4\phn}$ \\    
WSB 60            & 1.67 $^{+0.04}_{-0.03}$ & -0.60 $^{+0.02}_{-0.02}$
                  & \phn$7.28\pm0.06$ & 137 $^{+1\phn\phn}_{-1\phn\phn}$ 
                  & 0.24 $^{+0.05}_{-0.05}$ & 12 $^{+2\phn}_{-2\phn}$ \\    
SR 13             & 1.67 $^{+0.02}_{-0.04}$ & -0.83 $^{+0.01}_{-0.01}$
                  & \phn$7.28\pm0.06$ & 137 $^{+1\phn\phn}_{-1\phn\phn}$ 
                  & 0.48 $^{+0.10}_{-0.09}$ & 13 $^{+3\phn}_{-2\phn}$ \\    
DoAr 44           & 1.79 $^{+0.03}_{-0.02}$ & -0.63 $^{+0.03}_{-0.02}$
                  & \phn$6.79\pm0.16$ & 147 $^{+4\phn\phn}_{-3\phn\phn}$ 
                  & 1.81 $^{+0.37}_{-0.37}$ & 17 $^{+3\phn}_{-3\phn}$ \\    
RX J1633.9$-$2422 & 1.72 $^{+0.01}_{-0.02}$ & -0.60 $^{+0.03}_{-0.02}$
                  & \phn$6.79\pm0.16$ & 147 $^{+4\phn\phn}_{-3\phn\phn}$ 
                  & 1.04 $^{+0.24}_{-0.20}$ & 16 $^{+3\phn}_{-3\phn}$ \\    
WaOph 6           & 1.99 $^{+0.13}_{-0.08}$ & -0.41 $^{+0.22}_{-0.12}$
                  & \phn$7.87\pm1.37$ & 128 $^{+41\phn}_{-14\phn}$ 
                  & 2.79 $^{+2.68}_{-0.54}$ & 20 $^{+5\phn}_{-4\phn}$ \\
AS 209            & 1.96 $^{+0.07}_{-0.04}$ & -0.30 $^{+0.13}_{-0.08}$
                  & \phn$7.87\pm0.91$ & 127 $^{+22\phn}_{-10\phn}$ 
                  & 1.49 $^{+0.75}_{-0.30}$ & 15 $^{+3\phn}_{-3\phn}$ \\      
HD 163296         & 2.06 $^{+0.04}_{-0.04}$ & \phd0.13 $^{+0.09}_{-0.08}$
                  & \phn$8.43\pm0.78$  & 118 $^{+15\phn}_{-8\phn\phn}$ 
                  & 33.1 $^{+13.0}_{-6.7\phn}$ & 28 $^{+6\phn}_{-5\phn}$ \\
\enddata
\tablecomments{Quoted values are the peaks of the posterior distributions, with uncertainties that represent the 68\%\ confidence interval.  Inferences of $R_{\rm eff}$, $L_{\rm mm}$, $\varpi$, and $d$ are described in Sect.~\ref{sec:convert}; $L_\ast$ and $\langle T_d \rangle$ are defined in Sect.~\ref{sec:discussion} (the latter in Eq.~\ref{eq:meantd}).}
\end{deluxetable*}


\end{document}